\documentclass{article}[12pt]
\usepackage{jheppub}
\usepackage[utf8]{inputenc}
\usepackage{amsmath}
\usepackage{amssymb}
\usepackage{amsfonts}
\usepackage{amsthm}
\usepackage{xcolor}
\usepackage{rotating}
\usepackage{array}
\usepackage{appendix}
\usepackage{bbold}

\def\b1{\bar 1}

\title{Defect Conformal Blocks from Appell Functions}

\author{Ilija Buri\' c}
\author{and Volker Schomerus}
\affiliation{DESY, Notkestra\ss e 85, D-22607 Hamburg, Germany}

\preprint{DESY 20-206}

\abstract{We develop a group theoretical formalism to study correlation 
functions in defect conformal field theory, with multiple insertions of 
bulk and defect fields. This formalism is applied to construct the defect 
conformal blocks for three-point functions of scalar fields. Starting from 
a configuration with one bulk and one defect field, for which the correlation 
function is determined by conformal symmetry, we explore two possibilities,  
adding either one additional defect or bulk field. In both cases it is possible 
to express the blocks in terms of classical hypergeometric functions, though 
the case of two bulk and one defect field requires Appell's function $F_4$.}

\begin{document}

\maketitle

\newpage
\section{Introduction and Summary}

Conformal field theories in dimensions higher than two have received a renewed attention in the last
two decades. This is partly due to their conjectured role in the description of quantum gravity through the
celebrated holographic duality and partly due to the remarkable success of constraining them using the
conformal bootstrap program.

Even though bootstrap techniques are usually applied to correlation functions of local fields, they extend
to setups in which non-local operators are inserted along a $p$-dimensional hyperplane, so-called {\it defect}
conformal field theories. The presence of such a $p$-dimensional defect reduces the symmetry group to
consist of those conformal transformations of $\mathbb{R}^d$ that preserve the $p$-dimensional subspace
$\mathbb{R}^p$ along which the defect is localised. These form the subgroup $G_{d,p} = SO(1,p+1) \times
SO(d-p)$ of the bulk conformal group $G_d = SO(1,d+1)$. Defects can arise as an impurity in a critical
system, the boundary of an experimental setup or an insertion of a (Wilson, 't Hooft-) line operator  etc.
As in an ordinary conformal field theory, correlation functions of local operators in the presence of a defect
must satisfy a set of consistency conditions. Defects actually introduce two types of new data. On the one hand, they
support new defect fields $\hat \varphi$ that can only be inserted at points $\hat x \in \mathbb{R}^p$ along
the defect. These fields close with respect to operator products and the associated coefficients $c_{\dots}$
satisfy essentially the same type of crossing equations as local bulk fields do, only augmented by the
presence of the symmetry group $SO(d-p)$ that describes rotations transverse to the defect. Defect fields
can carry so-called transverse spin, i.e. they can transform non-trivially under the action of transverse
rotations just like fields in a bulk theory may transform under some internal symmetry. In addition to the
defect field operator product coefficients $c_{\dots}$ there is another set of new data a defect brings
in: the bulk-defect operator product coefficients $b_{..}$. The latter appear when a local bulk field is
moved close to the defect and expanded in terms of defect fields. Consistency constraints on the latter
arise from considering mixed correlators of bulk and defect fields.

In order to formulate these constraints, one needs to have a good control over the functional form of
correlators \cite{Cardy:1991tv,McAvity:1993ue,Lauria:2017wav}, possible tensor structures appearing in
them \cite{Kobayashi:2018okw,Guha:2018snh,Lauria:2018klo} and conformal blocks \cite{McAvity:1995zd,
Liendo:2012hy,Billo:2016cpy,Gadde:2016fbj,Isachenkov:2018pef}. Once all these ingredients are known,
the crossing equations are still difficult to study numerically because of the lack of positivity of
coefficients that enter into them (see however \cite{Gliozzi:2015qsa,Gliozzi:2016cmg}). However, the
equations may be amenable to analytic treatments, see e.g. \cite{Lemos:2017vnx,Liendo:2019jpu}. One of
the principal reasons to study defects is that they appear as tractable sectors of higher-dimensional
theories.  This is particularly the case for boundaries, in the presence of which two-point functions
of bulk operators depend on only one conformal invariant. Therefore, techniques of one-dimensional
conformal field theories can teach us about higher-dimensional theories \cite{Hogervorst:2017kbj,Bissi:2018mcq,
Kaviraj:2018tfd,Mazac:2018biw}.

In ordinary bulk conformal field theory, the functional form of a three-point function
is still fixed by conformal symmetry up to multiplicative constants. Crossing symmetry
constraints arise by considering the first case of non-trivial correlators with
$N=4$ local field insertions. In the presence of a defect, the analogue of the three-point
function is given by a two-point function of a local bulk and a local defect field. In fact,
such correlators are well known to be determined by the residual conformal symmetry $G_{d,p}$
up to multiplicative constants. The first examples of correlators with interesting dependence
on the insertion points arise when we add one more field to this minimal configuration. This
can be either a defect field or a bulk field. The associated conformal partial wave expansions
and conformal blocks have not received much attention yet, except for the special case of two
bulk field insertions with a trivial (identity) defect field, see \cite{McAvity:1995zd,Liendo:2012hy,Billo:2016cpy}.
Results on setups with non-trivial defect field are rare - see \cite{Lauria:2020emq}, however, where
blocks for two defect and one bulk fields were constructed recently. This work also contains some
results on the more difficult case of two bulk and one defect field. But in determining the
dependence on only two of the three cross ratios that characterise such configurations,
\cite{Lauria:2020emq} falls short of identifying these blocks.
\medskip

The main new result of this work is the construction of conformal partial waves for three-point functions of
two scalar bulk and one scalar defect field in terms of Appell functions. To be more specific, let
us consider two scalar bulk fields $\varphi_i(x_i)$ of scaling weights $\Delta_i$ inserted at two points $x_i$ of
the $d$-dimensional bulk space. The defect field is denoted by $\hat \varphi_3(\hat x_3)$. It has scaling
weight $\Delta_{\hat 3}$ and is inserted at a point $\hat x_3$ on the $p$-dimensional defect. It will be useful to split the
coordinates of the bulk field insertion points as $x_i = (\hat x_i,x_{i\perp})$ into the first $p$ components that
parametrise a position along the defect and the remaining $q=d-p$ components for the transverse position. The conformal
transformations along the defect allow us to move $\hat x_3, \hat x_i$ to the points $\hat x_1 = 0$, $\hat x_2 = 1 e_1$ and
$\hat x_3 = \infty$. The resulting configuration is left invariant by the subgroup $SO(p-1) \times SO(d-p) \subset G_{d,p}$.
We can use rotations in transverse space to rotate $x_{1\perp}$ into one direction of transverse space and $x_{2\perp}$ into
a plane that is spanned by a second direction, provided the co-dimension $q = d-p \geq 2$. Hence, the configuration of the
three insertion points is characterised by three cross ratios. In the frame we just described, these are the two distances
of the bulk field insertions from the defect and the angle between the insertions. Before we fix the frame the cross ratios
read
\begin{equation}
\cos\kappa = \frac{x_{1\perp}\cdot x_{2\perp}}{|x_{1\perp}| |x_{2\perp}|} \ , \quad
v_i  =  - x^2_{i^\vee\perp}\frac{x_{i 3}^4}{\hat x_{12}^2 x_{13}^2 x_{23}^2 +
(\hat x_{13}^2 x_{23}^2 - \hat x_{23}^2 x_{13}^2)(x_{23}^2 - x_{13}^2)}
\end{equation}
where $x^2_{i3} = x^2_{i\perp} + \hat x^2_{i3}$ for $i=1,2$ and we introduced $i^\vee = 3-i$. Note that with the gauge
choice we specified above, $v_i = - x^2_{i^\vee\perp}$ and $\kappa$ is indeed the angle between $x_{i\perp}$. Having
constructed all relevant coordinates we can now spell out the conformal partial wave expansion of the three-point
function under consideration,
\begin{equation}
\langle \varphi_1(x_1) \varphi_2(x_2) \hat \varphi_3(\hat x_3) \rangle =
 \frac{1}{|x_{1\perp}|^{\Delta_1} |x_{2\perp}|^{\Delta_2}} \left(\frac{|x_{2\perp}|}{x_{23}^2}\right)^{\Delta_{\hat 3}}\,
 \sum_{\hat \Delta,\hat \Delta',s} \lambda^{123}_{\hat \Delta,\hat \Delta',s} \psi_{\hat \Delta,\hat \Delta',s}(v_1,v_2,\kappa)
\end{equation}
where the blocks are given by
 \begin{equation}\label{intro-1}
    \psi_{\hat\Delta,\hat\Delta',s}(v_1,v_2,\kappa) = v_1^{\frac{\hat\Delta}{2}-\frac{\Delta_{\hat3}}{4}} v_2^{\frac{\hat\Delta'}{2}-\frac{\Delta_{\hat3}}{4}} \left(\frac{v_2}{v_1}\right)^{\frac{\Delta_{\hat 3}}{4}} \,
    F(v_1,v_2) \ C^{(d-p-2)/2}_s (\cos\kappa)\ .
\end{equation}
Functions $C^{(d-p-2)/2}_s$ are Gegenbauer polynomials and $F$ is given in terms of Appell's hypergeometric function as
\begin{equation}\label{intro-2}
    F(v_1,v_2) = F_4\left(\frac{\hat\Delta+\hat\Delta'-\Delta_{\hat3}}{2},\frac{\hat\Delta+\hat\Delta'-\Delta_{\hat3}+2-p}{2},
    \hat\Delta-\frac{p}{2}+1,\hat\Delta'-\frac{p}{2}+1;v_1,v_2\right)\ .
\end{equation}
The summation runs over scalar defect primaries $\hat \varphi$ and $\hat \varphi'$ or weight $\hat \Delta$ and $\hat \Delta'$,
respectively, that appear in the bulk-defect operator product of the two bulk fields. The label $s$ denotes the transverse
spin of $\varphi,\varphi'$. Note that we assumed the transverse spin of the defect field $\varphi_3$ to vanish so that for the
three-point function to be non-zero, the two defect fields $\hat \varphi$ and $\hat\varphi'$ must have the same transverse spin
$s$. The appearance of the Gegenbauer polynomials for the two-point function of spinning fields in transverse space is standard.
What is really new about our result is the construction of the block in the variables $v_i$, at least when the defect field
$\hat\varphi_3$ is non-trivial. The case of trivial $\hat\varphi_3$ corresponds to a two-point function of two scalar bulk fields. The
associated blocks were first constructed in \cite{McAvity:1995zd} for $p = d-1$ (boundary) and then for more general defects
in \cite{Billo:2016cpy}. In addition, the blocks of the three-point functions with non-trivial defect field have been studied
quite recently in \cite{Lauria:2020emq,Behan:2020nsf}. The authors of \cite{Lauria:2020emq} found that along the diagonal $v_1 = v_2$ the blocks could
be expressed in terms of $_4F_3$. In \cite{Behan:2020nsf}, the blocks for a general configuration of points were computed (in the case of the boundary defect), however for very special quantum numbers of intermediate fields that are allowed to propagate when the bulk field is free. These partial waves are expressed in terms of Gauss' hypergeometric function. Here we reproduce both of these results by taking appropriate special cases of the Appell function.

\medskip

In order to obtain these results we further develop the group theoretic approach to conformal blocks in general
and defect blocks in particular, see e.g. \cite{Schomerus:2016epl,Isachenkov:2018pef,Buric:2019dfk,
Buric:2020buk}. The main strategy of this approach is to uplift correlation functions in the (defect) conformal field
theory to an appropriate  space of functions on the conformal group (or several copies thereof) and then to construct
the blocks as eigenfunction of the Laplace-Beltrami operator, thereby making a direct link to harmonic analysis. The
relevant restrictions of the Laplace-Beltrami operators take the form of Schroedinger operators for various integrable
multi-particle quantum mechanical models of Calogero-Sutherland type which have often been studied extensively in the
mathematical literature. The starting point of the uplift to the conformal group are realisations of the space of
field insertion points through quotients of the conformal symmetry group. In the presence of defects one needs to
distinguish two symmetry groups, namely the conformal symmetry $G_d = SO(1,d+1)$ of the bulk and the subgroup $G_{d,p}
= SO(1,p+1) \times SO(d-p)$ that is preserved by the defect. While previous work on group theoretic constructions of
defect blocks has focused on uplifts to the bulk symmetry $G_d$ and hence on what is known as bulk channel blocks, the
present paper is the first to consider uplifts to the defect conformal group $G_{d,p}$ and hence on what is known as
the defect channel. This also allows us to include correlators involving defect fields $\hat \varphi$ for the first time,
such as the one discussed above. While the uplift of defect fields to the defect conformal group $G_{d,p}$ follows
very closely the corresponding constructions of bulk-field uplifts in bulk conformal field theory, see in particular
\cite{Buric:2019dfk}, the uplift of bulk fields to $G_{d,p}$ is more challenging. In this work we construct this
uplift explicitly for arbitrary scalar bulk fields and any pair of dimensions $d,p$, see eqs.\ \eqref{bulk-prefactor},
\eqref{bulk-lift} (with $\mu = 1$), \eqref{choice-of-embedding} and \eqref{scalar-phi}. Once the uplift of the
individual fields is understood, it is straightforward to uplift correlations functions of any numbers $m$ and $n$
of bulk- and defect-field insertions. In general, such an uplift will be described through functions on a product
of $n+m -1$ copies of the defect conformal group $G_{d,p}$. It turns out, however, that it is possible to reduce
the number of factors by forming pairs of bulk and defect fields. The process allows to uplift correlation
functions of at least up to two bulk and two defect fields to functions on a single copy of the defect conformal group,
just as it is the case for four-point functions of bulk fields in ordinary bulk conformal field theory.

Let us now briefly outline the content of the next sections. Section 2 is devoted to the uplift of bulk
and defect fields to the defect conformal group. In particular the discussion of bulk field uplifts is
entirely new. While the general discussion applies to spinning fields as well, the concrete solution
that is offered is restricted to scalar bulk fields. In section 3 we uplift correlators with any numbers
of bulk- and defect-field insertions and discuss how one can reduce the number of factors by pairing
bulk and defect fields. As a simple application we compute the two-point function of one defect and
one bulk field for any pair of dimensions $d$ and $p$, including defect fields with spin. Section 4
finally addresses correlators that involve non-trivial cross-ratios and hence conformal blocks. As a
warmup we construct and solve the Casimir equations for correlation functions of two bulk fields
(without any defect field) in section 4.2. This setup is well studied and we reproduce known
results. Next we turn to correlators of two defect fields and a single bulk field in subsection 4.2.
The setup only admits a single cross ratio. Once again we construct and solve the relevant Casimir
equation. The results slightly extend recent work on such correlators in \cite{Lauria:2020emq}. Our
results on correlation functions of two bulk and one defect field, see the discussion above, are obtained
in subsection 4.3. The paper concludes with a short summary and a list of open problems.

\newpage
\subsection*{Glossary and notations}

For the reader's convenience let us briefly collect a few notations, many of which we have introduced
already while explaining background, results and methods of this work. Throughout the paper we will
consider a $p$-dimensional defect in $d$-dimensional spacetime and use the notation
\begin{equation}
    M = S^d = \mathbb{R}^d\cup\{\infty\}, \quad N = S^p = \mathbb{R}^p\cup\{\infty\}\ .
\end{equation}
We will often ignore the difference between the vector spaces $\mathbb{R}^d, \mathbb{R}^p$ and their conformal
compactifications $M,N$ as the distinction has no bearing on the questions that will be considered. A generic
spacetime point will be written as $x$, and a point on the defect as $\hat x$. Components of $x$ parallel and
orthogonal to the defect are denoted $x_\parallel=\hat x$ and $x_\perp$, respectively. A standard orthonormal
basis for $\mathbb{R}^d$ is written as $\{e_\mu\}$, with $\{e_a\}$, $a=1,...,p$ spanning the defect subspace.

The group of conformal transformations of $M$ is denoted $G_d = SO(d+1,1)$. Its subgroup of transformations
that map $N$ to itself, the defect conformal group, is $G_{d,p}=SO(p+1,1)\times SO(q)$. Here we use the standard
notation $q=d-p$. The $SO(p+1,1)$ factor is generated by dilations $D$, translations and special conformal
transformations along the defect $P_a,K_a$, and rotations in the defect plane $M_{ab}$, while the $SO(q)$
factor is generated by transverse rotations $M_{ij}$, $i,j = p+1,...,d$. For any element $g$ of the defect
conformal group, $g_p$ and $q_q$ will stand for its unique factors in $SO(p+1,1)$ and $SO(q)$.

In some of the formulas, a distinction has to be made between two cases: $q>1$ and $q=1$. As written, our
results will apply to the former case. Modifications for $q=1$ are mostly evident.

\section{Lifting Conformal Primary Fields}

In this section, we will show how conformal primary fields, both those in the bulk and the ones on the defect,
can be uplifted to functions on the symmetry group $G_{d,p}$ of the defect theory. By construction, the uplift 
will respect the usual action of conformal transformations on primary fields. As a warmup, we shall first address 
the defect fields for which the construction is analogous to the case of bulk fields in ordinary conformal field 
theory, see \cite{Buric:2019dfk}. The rest of the section is then devoted to lifting bulk primary fields to 
the defect conformal group. After an outline of the general strategy, we briefly recall the Iwasawa decomposition 
of the conformal group, which is the principal technical tool in lifting bulk fields. The lift is finally 
constructed explicitly for scalar bulk fields.  

\subsection{Lifting defect primaries}

Fields $\hat \varphi$ that can be inserted at points $\hat x$ along the $p$-dimensional defect may be lifted 
to the defect conformal group $G_{d,p}$ very much in the same way as one lifts bulk fields to the full conformal 
group, \cite{Buric:2019dfk}. Here we will review this construction with an eye on our subsequent discussion 
of bulk fields and their lift to the defect group. 

\vskip0.1cm Defect primary fields $\hat \varphi: N \xrightarrow{} \hat W$ can take values in a finite dimensional 
representation $\hat \rho$ of the subgroup $\hat K_{d,p} = SO(1,1) \times SO(p) \times SO(q) \subset 
G_{d,p}$ of the defect conformal group. The choice of representation is used to characterise the weight, 
spin and so-called transverse spin of the defect field. The action of $\hat K_{d,p}$ on $\hat W$ gives rise
to a representation $\hat\pi$ of the defect conformal group on the defect primary,
\begin{equation}\label{representation-on-a-field-defect}
    (\hat\pi_h\hat \varphi)(h \hat x) = \hat \rho(dh_{\hat x})\hat \varphi(\hat x)\ .  
\end{equation}
Here, $h$ is an arbitrary element of $G_{d,p}$ and $dh$ is its differential when $h$ is considered as 
a diffeomorphism of $M$. Often this action is written in terms of infinitesimal transformations which 
act on primary fields as first order differential operators depending on weight, spin and transverse
spin.  

\vskip0.1cm To begin our discussion of the lift to $G_{d,p}$ we note that the action of the defect conformal group 
on the points $\hat x$ is transitive. Therefore, the space $N$ of defect-field insertion points can be 
represented as a quotient $N = G_{d,p}/\hat S_{d,p}$ where $\hat S_{d,p}$ denotes the stabiliser
subgroup of a point on the defect. Up to conjugation with an element of $G_{d,p}$, this stabiliser 
subgroup is unique and is isomorphic to the group
\begin{equation}    
\hat S_{d,p} = \left(SO(1,1)\times SO(p)\right) \ltimes \mathbb{R}^p\times SO(q),
\end{equation} 
generated by dilations, special conformal transformations and (parallel and transverse) rotations. The group $\hat S_{d,p}$ acts on the defect conformal group $G_{d,p}$ through right multiplication and with 
respect to this action, $G_{d,p}$ decomposes into orbits. In order to lift a defect field $\hat \varphi$ 
from $N$ to the defect conformal group we must fix an embedding   
\begin{equation}
    g_p : N \xrightarrow{} G_{d,p},
\end{equation}
that intersects almost all\footnote{We use the phrase "almost all" in the measure-theoretic sense, i.e. it 
means "up to sets of measure zero".} orbits of the right action of $\hat S_{d,p}$ on $G_{d,p}$ exactly once. 
The embedding we will be using is given by 
\begin{equation}\label{subgroup-embedding-defect}
     g_p(\hat x) = e^{\hat x^a P_a}  \ .
\end{equation}
Let us note that it is indeed possible to factorise almost all elements $g \in G_{d,p}$ as $g = 
g_p(\hat x) \hat s$ for some point $\hat x \in N$ and some element $\hat s \in \hat S_{d,p}$. 
Consequently, the factorisation of the product 
\begin{equation}
    h g_p(\hat x) = g_p(\hat y(\hat x,h)) \hat s_p(\hat x,h)\quad \textit{ for } \ h \in G_{d,p}, 
\end{equation}
defines two functions $\hat y(\hat x,h)$ and $\hat s_p(\hat x,h)$. These are essentially the same as 
in a bulk conformal field theory \cite{Dobrev:1977qv,Buric:2020buk}, since transverse rotations commute with the image of $g_p$.
The first function $\hat y(\hat x,h)$ gives rise to an action of the defect conformal group on $N$ that is easily seen to coincide with the usual geometric action of 
$G_{d,p}$ on points $\hat x$, i.e. $\hat y(\hat x,h) = h \hat x$.
  
As the lift of the defect primary $\hat \varphi$ we define the function $\hat f:G_{d,p} \xrightarrow{} \hat W$ which agrees with $\hat\varphi$ on $g_p(N)$
\begin{equation}\label{defect-lift}
    \hat f(g_p (\hat x)) = \hat \varphi(\hat x),
\end{equation}
and transforms covariantly under right regular transformations with $\hat s \in \hat S_{d,p}$, 
\begin{equation} \label{hatfcovariance}
   \hat f(g \hat s) = \hat\mu(\hat s)^{-1} \hat f(g),\quad g\in G_{d,p},\ \hat s\in \hat S_{d,p}\ .
\end{equation}
Here, $\hat\mu$ is a finite dimensional representation of $\hat S_{d,p}$ obtained from the representation $\hat\rho$ by trivial extension to the abelian factor $\mathbb{R}^p$. Clearly, conditions $(\ref{defect-lift})$ and $(\ref{hatfcovariance})$ define $f$ uniquely almost everywhere on $G_{d,p}$. It is a little less obvious, but still not difficult to see that under the lift $\hat\varphi\mapsto \hat f$, the action of defect conformal transformations $(\ref{representation-on-a-field-defect})$ is carried to the left-regular action on $\hat f$ - for a proof see \cite{Dobrev:1977qv,Buric:2020buk}.

\subsection{Lifting bulk primaries}

Our main goal now is to lift bulk primary fields $\varphi$ to the defect conformal group in a 
sense similar to what we did in the previous subsection for defect fields. In this case, however, 
the construction is entirely new and not quite as straightforward.   

\subsubsection{Geometric lifts and intertwiners}

In order to extend a bulk field $\varphi : M\xrightarrow{}W$ which takes values in some vector
space $W$ to a vector valued function $f : G_{d,p}\xrightarrow{}W$ on the defect conformal group we
need to specify four pieces of data. First, we need to specify an embedding of the bulk space $M$ into 
the defect conformal group, 
\begin{equation}
    g_d : M \xrightarrow{} G_{d,p}\ .
\end{equation}
On the submanifold $g_d(M)$ we require that $f$ agrees with $\varphi$, possibly up to some specified
prefactor $\Phi(x)$
\begin{equation}\label{bulk-prefactor}
    f(g_d(x)) = \Phi(x) \varphi(x)\ .
\end{equation}
Next, we pick a subgroup $S_{d,p}\subset G_{d,p}$ such that almost all of its orbits in $G_{d,p}$ (under
the right regular action) intersect $g_d(M)$ exactly once. We postulate that $f$ is right-covariant with
respect to $S_{d,p}$
\begin{equation}\label{bulk-lift}
    f(g s) = \mu(s)^{-1} f(g),\quad g\in G_{d,p},\ s\in S_{d,p}\ .
\end{equation}
Here, $\mu:S_{d,p}\xrightarrow{}\text{Aut}(W)$ is some finite-dimensional representation of $S_{d,p}$. 
Clearly, the function $f$ is uniquely determined almost everywhere on $G_{d,p}$ by the properties 
\eqref{bulk-prefactor} and \eqref{bulk-lift}. We shall call the quadruple $(g_d,S_{d,p},\mu,\Phi)$ 
a lift of a bulk field. 
\smallskip

The data just introduced allows to factorise almost all elements of $G_{d,p}$ uniquely as $g = g_d(x)
s$. Moreover, it defines an action of $G_{d,p}$ on the space $M$ as follows. Given any $h\in G_{d,p}$, the
factorisation
\begin{equation}
    h g_d (x) = g_d(y(x,h)) s_d(x,h),
\end{equation}
defines functions $y(x,h)$ and $s_d(x,h)$. In particular, the function $y(x,h)$ is an action of the
group $G_{d,p}$ on the bulk space. Indeed, we can write $h_1 h_2 g_d(x)$ in two ways
\begin{align*}
   & h_1 h_2 g_d (x) = g_d(y(x,h_1 h_2)) s_d(x, h_1 h_2)\\
   & = h_1 g_d(y(x,h_2)) s_d(x,h_2) = g_d (y(y(x,h_2),h_1)) s_d(y(x,h_2),h_1) s_d(x,h_1)\ .
\end{align*}
The product of last two terms in the second line is again an element of $S_{d,p}$ so that we can 
conclude
\begin{equation*}
    y(x,h_1 h_2) = y(y(x,h_2),h_1)\ .
\end{equation*}
This precisely says that $y(x,h)$ is an action of $G_{d,p}$ on $M$. In principle, this action may or may
not coincide with the geometric action of the defect conformal group on the bulk.\footnote{The geometric
action of a transformation $h$ on the point $x$ is always written as $hx$.} This depends both on the map $g_d$ and
on the subgroup $S_{d,p}$, but not on $\Phi$ or $\mu$. If the action $y(x,h)$ is the geometric action on
$M$ we shall say that the lift is geometric.

\vskip0.1cm The space of fields $\varphi: M \xrightarrow{} W$ carries a representation of the conformal group $G_d$, and thus by restriction, of the defect group $G_{d,p}$ as well. Similarly as above, the representation,
denoted $\pi$, is given by
\begin{equation}
    (\pi_h\varphi)(h x) =  \rho(dh_x)\varphi(x)\ .
\end{equation}
Here, $\rho$ is the representation of the group $K_d=SO(1,1)\times SO(d)$ of dilations and rotations that characterises transformation properties of $\varphi$, its conformal weight and its spin. As before,
$h$ is an arbitrary element of $G_{d,p}$ and $dh$ is its differential when $h$ is considered as a smooth map $M\xrightarrow{}M$. Thanks to the identity $dh_x = k(x,h)$, we can rewrite the last equation as
\begin{equation}\label{representation-on-a-field}
(\pi_h\varphi)(h x) = \rho(k(x,h))\varphi(x)\ .
\end{equation}
The element $k(x,h)$ is the factor appearing in the Bruhat decomposition of $G_d$
\begin{equation}\label{Bruhat}
    h m (x) = m(hx)\ n(z(x,h))\ k(x,h)\ .
\end{equation}
Let us give more details. In the conformal group $G_d$ almost any element can be written as $g = m n k$, with
\begin{equation}
    m(x^\mu) = e^{x^\mu P_\mu}, \quad n_\nu(x^\mu) = w^{-1}_\nu m(x^\mu) w_\nu, \quad k\in K_d = SO(1,1)\times SO(d)\ .
\end{equation}
Here $w$ is the Weyl inversion. It is defined as any one of the following group elements
\begin{equation}
    w_\mu = e^{\pi\frac{K_\mu - P_\mu}{2}}\ .
\end{equation}
The element $w_\mu$ acts on the space $\mathbb{R}^d$ as the composition of the conformal inversion
$I$ and the reflection $s_{e_\mu}$ in the hyperplane orthogonal to the basis vector $e_\mu$. The function $z(x,h)$ 
depends on the choice of $\mu$, but of course the corresponding factor in eq.\ \eqref{Bruhat} remains the same. For more details on the Bruhat decomposition and its relation to Ward identities in ordinary conformal field theory, the reader is
referred to \cite{Buric:2019dfk,Buric:2020buk,Dobrev:1977qv}.

The space of functions $f$ that have covariance properties \eqref{bulk-lift}  also carries a
representation of the defect conformal group obtained by the restriction of the left regular
representation. The space of functions with this action of $G_{d,p}$ is known as the induced
module $\text{Ind}_{S_{d,p}}^{G_{d,p}}\mu$. We wish to understand under which conditions on $\Phi$ and
$\mu$ a geometric lift $(g_d,S_{d,p},\mu,\Phi)$ is actually an intertwiner between the above
two modules. Let $f_\varphi$ be the lift of a function $\varphi$ and denote by $L$ the left
regular representation of $G_{d,p}$. Then we have
\begin{align*}
      L_{h^{-1}}f_\varphi(g) &= f_\varphi(h g) = f_\varphi(h g_d(x) s) = f_\varphi\left(g_d(y(x,h)) s_d(x,h) s\right) =\mu(s^{-1}) \mu(s_d(x,h)^{-1}) \Phi(y(x,h))) \varphi(y(x,h))\\
     & = \mu(s^{-1}) \Phi(x) \pi_{h^{-1}}\varphi(x)   = \mu(s^{-1}) f_{\pi_{h^{-1}}\varphi} (g_d(x)) =  f_{\pi_{h^{-1}}\varphi}(g)\ .
\end{align*}
In order to get to the second line we have used the fact that the lifts are geometric and the identity
\begin{align*}
  \mu(s_d(x,h)^{-1}) \Phi(hx) \varphi(hx) = \Phi(x) \pi_{h^{-1}}\varphi(x)\ .
\end{align*}
Looking back at eq.\ \eqref{representation-on-a-field} , a sufficient condition for this identity to hold is
\begin{equation}\label{eqn-mu-Phi}
    \Phi(hx)^{-1}\mu(s_d(x,h))\Phi(x) = \rho(k(x,h)) \ .
\end{equation}
This is an equation for both $\Phi$ and $\mu$ that were left completely arbitrary by the requirement
of geometricity of the lift. If they are satisfied, the lift is an intertwiner between the representation
on fields $\varphi$ and the left regular representation of $G_{d,p}$ restricted to the space of right
$S$-covariant functions.

\subsubsection{The Iwasawa decomposition}

Before we address the construction of a geometric lift of bulk fields, we need to introduce our
principal mathematical tool, the Iwasawa decomposition of the group $G_p = SO(p+1,1)$. The term
``Iwasawa decomposition'', \cite{Dobrev:1977qv}, can refer to either of the following two ways
to factorise elements of $G_p$
\begin{equation}\label{Iwasawa}
    G_p = N_I A_I K_I, \quad G_p = \tilde N_I A_I K_I\ .
\end{equation}
We will be interested in both of these decompositions. Here, the notation is
\begin{equation}
    K_I = SO(p+1), \quad A_I = SO(1,1), \quad N_I = e^{\mathfrak{g}_{-1}}, \quad \tilde N_I = e^{\mathfrak{g}_1}.
\end{equation}
That is, $\tilde N_I$ is the group of translations and $N_I$ that of special conformal transformations.
The other factors are the $1$-dimensional group of dilations $A_I$ and the maximal compact subgroup
$K_I$ of $G_p$, generated by rotations and differences $P_a - K_a$ of translation and special conformal
generators. Another standard decomposition of $G_p$ that plays a prominent role in our constructions is
the Bruhat decomposition that we reviewed above. It is important to understand how these two
decompositions are related. If we, for definiteness, consider the second of factorisations \eqref{Iwasawa} ,
the relation will be known as soon as we find the Iwasawa factors of $g = e^{x^a K_a}$. Elements of $G_p$
which are Bruhat factors of other types, that is, translations, rotations and dilations are by themselves
Iwasawa factors as well. In the appendix A, it is shown that
\begin{equation}\label{Iwasawa-ek}
    e^{x^a K_a} = e^{\frac{x^a}{1+x^2}P_a}(1+x^2)^{-D} k_I(x)\ .
\end{equation}
Here, the last factor $k_I(x)$ reads in the $p+2$-dimensional representation of $G_p$
\begin{equation}\label{KI}
    k_I(x) = \begin{pmatrix}
    1 & 0 & 0\\
    0 & \frac{1-x^2}{1+x^2} & \frac{-2x^a}{1+x^2}\\
    0 & \frac{2 x^a}{1+x^2} & \delta_{ab} - \frac{2x^a x^b}{1+x^2}
    \end{pmatrix}\ .
\end{equation}
We have written the matrix on the right hand side in the block form, as indicated by indices carried by
matrix elements. The usefulness of the formula \eqref{Iwasawa-ek}  will be seen in the following
subsections. One can conjugate both sides by $w_p$ to obtain another variant of it, that we will use frequently
\begin{equation}\label{Iwasawa-2}
    e^{x^a P_a} = e^{\frac{x^a}{1+x^2}K_a}(1+x^2)^{D} k_I(-x)\ .
\end{equation}
It is possible to read equations \eqref{Iwasawa-ek}  and \eqref{Iwasawa-2}  as Bruhat decompositions of $k_I(x)$ as well
\begin{equation}\label{Bruhat-of-kI}
    k_I(x) = e^{-x^a P_a} e^{\frac{x^a}{1+x^2}K_a}(1+x^2)^D = (1+x^2)^{-D} e^{\frac{x^a}{1+x^2}K_a} e^{-x^a P_a}\ .
\end{equation}

\subsubsection{Construction of the lift}

Let us now explicitly construct a geometric lift of bulk fields to the defect conformal group. The group with respect to which the associated function $f$ is required to be covariant is
\begin{equation}\label{group-S}
        S_{d,p} = SO(p+1) \times SO(q-1)\ .
\end{equation}
The group $S_{d,p}$ is generated by rotations in the defect plane, transverse rotations that preserve one particular direction, say $e_d$, together with elements of the form $P_a - K_a$. We turn to the embedding of the bulk space into the group, $g_d : \mathbb{R}^d\xrightarrow{}G_{d,p}$. Let us set
\begin{equation}\label{choice-of-embedding}
    g_d(x) = e^{x^a P_a} |x_\perp|^D e^{\varphi^i M_{id}}\ .
\end{equation}
Here $\varphi^{p+1},...,\varphi^{d-1}$ are the angles of a spherical coordinate system on $\mathbb{R}^q$. To be precise, these coordinates are defined in such a way that $e^{\varphi^i M_{id}}$ maps the vector $e_d$ to $x_\perp/|x_\perp|$ in $\mathbb{R}^q$. There is a unique element with this property of the above form (see appendix B). To show that the pair $(S_{d,p},g_d)$ defines a geometric lift, we determine $y(x,h)$ and $s_d(x,h)$
\begin{align*}
    & m(\hat x') e^{x^a P_a} |x_\perp|^D e^{\varphi^i M_{id}} = e^{(\hat x'+x)^a P_a} |x_\perp|^D e^{\varphi^i M_{id}}\ \ \implies\ \ y(x,m(\hat x')) = m(\hat x')x, \quad s_d(x,m(\hat x')) = 1,\\
    & e^{\lambda D} e^{x^a P_a} |x_\perp|^D e^{\varphi^i M_{id}} = e^{(e^{\lambda} x)^a P_a} (e^\lambda|x_\perp|)^D e^{\varphi^i M_{id}}\ \ \implies\ \ y(x,e^{\lambda D}) = e^{\lambda D}x, \quad s_d(x,e^{\lambda D})=1,\\
    & r_p e^{x^a P_a} |x_\perp|^D e^{\varphi^i M_{id}} = e^{(r_p x)^a P_a} |x_\perp|^D e^{\varphi^i M_{id}} r_p\ \ \implies\ \ y(x,r_p) = r_p x,\quad s_d(x,r_p)=r_p\ .
\end{align*}
Here $r_p\in SO(p)$. Let us now consider transverse rotations $r_q\in SO(q)$. In order for the decomposition
\begin{equation}\label{rq-decomposition}
    r_q e^{x^a P_a} |x_\perp|^D e^{\varphi^i M_{id}} = e^{x^a P_a} |x_\perp|^D r_q e^{\varphi^i M_{id}} =  e^{x^a P_a} |x_\perp|^D e^{\psi^i M_{id}} r'_{q-1},
\end{equation}
to hold, we must have the equality
\begin{equation}
    r_q e^{\varphi^i M_{id}} = e^{\psi^i M_{id}} r_{q-1}'\ .
\end{equation}
Here the factor $r'_{q-1}$ belongs to the group $SO(q-1)$ that stabilises the vector $e_d$. If one acts with both sides on $e_d$, one learns
\begin{equation}
    \frac{1}{|x_\perp|} r_q(x_\perp) = r_q\left(\frac{x_\perp}{|x_\perp|}\right) = e^{\psi^i M_{id}}(e_d) = \frac{y_{\perp}(x,r_q)}{|y_\perp(x,r_q)|}\ .
\end{equation}
But we already know from the dilation factor in the decomposition \eqref{rq-decomposition}  that $|y_\perp(x,r_q)| = |x_\perp|$, so we can conclude
\begin{equation}
    y(x,r_q) = r_{q}x, \quad s_d(r_q,x) = r_{q-1}'\ .
\end{equation}
The precise form of $r_{q-1}'$ is not important for us at the moment, but we observe that $r_{q-1}' = r_q$ whenever $r_q\in SO(q-1)$. This follows from the fact that the space spanned by $\{M_{id}\}$ is closed under conjugation by elements in $SO(q-1)$ (it carries the vector representation under the adjoint action). Finally, the action of the Weyl inversion $w_p$ is found with the help of the Iwasawa decomposition 
\begin{align*}
      w_p g_d(x) & = e^{\left(\frac{s_p x_\parallel}{|x_\parallel|^2}\right)^a P_a} e^{-s_p x_\parallel^a K_a} |x_\parallel|^{-2D} s_{e_p}s_{x_\parallel} |x_\perp|^D e^{\varphi^i M_{id}}\\
     & = e^{\left(\frac{s_p x_\parallel}{|x_\parallel|^2}\right)^a P_a}\left(\frac{|x_\perp|}{|x_\parallel|^{2}}\right)^D e^{-\frac{|x_\perp|}{|x_\parallel|^2}s_p x_\parallel^a K_a} s_{e_p}s_{x_\parallel}e^{\varphi^i M_{id}} \\
     & = e^{\left(\frac{s_p x_\parallel}{|x_\parallel|^2}\right)^a P_a}\left(\frac{|x_\perp|}{|x_\parallel|^{2}}\right)^D e^{\left(-\frac{|x_\perp|s_p x_\parallel/|x_\parallel|^2}{1+|x_\perp|^2/|x_\parallel|^2}\right)^a P_a} \left(1+\frac{|x_\perp|^2}{|x_\parallel|^2}\right)^{-D} k_I\left(\frac{-|x_\perp|s_p x_\parallel}{|x_\parallel|^2}\right) s_{e_p}s_{x_\parallel} e^{\varphi^i M_{id}}\\
     & = e^{\left(\frac{s_p x_\parallel}{|x|^2}\right)^a P_a}\left(\frac{|x_\perp|}{|x|^{2}}\right)^D e^{\varphi^i M_{id}} k_I\left(\frac{-|x_\perp|s_p x_\parallel}{|x_\parallel|^2}\right) s_{e_p}s_{x_\parallel}\ .
\end{align*}
The decomposition was used to get to the second to last line by an application of eq.\ \eqref{Iwasawa-ek} . We also applied the $G_p$-Bruhat decomposition in the first step. Other manipulations in the derivation above, like moving dilations past rotations and special conformal transformations, are evident. We read off
\begin{equation}
    y(x,w_p) = w_px, \quad s_d(x,w_p) = k_I\left(\frac{-|x_\perp|s_p x_\parallel}{|x_\parallel|^2}\right) s_{e_p}s_{x_\parallel}\ .
\end{equation}
Elements of the form $m(\hat x'), e^{\lambda D},r_p,r_q$ together with the Weyl inversion $w_p$ generate the whole defect conformal group. Therefore, we have the following important corollary
\begin{equation}\label{remarkable}
    y(x,h) = h x, \quad h\in G_{d,p}\ .
\end{equation}
The action $y(x,h)$ defined in the manner explained above through the choice of the group \eqref{group-S}  and the embedding \eqref{choice-of-embedding}  is precisely the action of the defect conformal group on the bulk space. That is, $(g_d,S_{d,p})$
gives rise to a geometric lift. The equation \eqref{remarkable}  is the most important result of this work and all subsequent
applications will rely on it in an essential way.
\medskip

Having found the group $S_{d,p}$ and the map $g_d$, we still need to solve equations \eqref{eqn-mu-Phi}  in order to turn a lift into a morphism of representations. For various types of elements, the equations read
\begin{align}
    & \Phi(m(\hat x')x)^{-1}\Phi(x) = 1,\quad \Phi(e^{\lambda D}x)^{-1} \Phi(x) = e^{-\Delta\lambda}, \quad \Phi(r_p x)^{-1} \mu(r_p) \Phi(x) = \rho(r_p),\\
    & \Phi(r_{q}x)^{-1}\mu(r_{q-1}')\Phi(x) = \rho(r_q), \quad \Phi(w_p x)^{-1}\mu\left(k_I\left(\frac{-|x_\perp|s_p x_\parallel}{|x_\parallel|^2}\right)s_{e_p}s_{x_\parallel}\right)\Phi(x) = \rho(k(x,w_p))\ .
\end{align}
We learn that $\Phi$ is a homogeneous function of $x_\perp$, that is
\begin{equation}\label{first-conditions}
    \Phi =\Phi(x_\perp), \quad \Phi(\lambda x_\perp) = \lambda^{\Delta}\Phi(x_\perp)\ .
\end{equation}
Under these conditions, the remaining equations simplify
\begin{align}\label{other-conditions-1}
    & \Phi(x_\perp)\rho(r_p) = \mu(r_p)\Phi(x_\perp), \quad \Phi(r_q x_\perp)\rho(r_{q})  = \mu(r_{q-1}')\Phi(x_\perp),\\
    & \Phi(x_\perp)^{-1} \mu\left(k_I\left(\frac{-|x_\perp|s_p x_\parallel}{|x_\parallel|^2}\right)s_{e_p}s_{x_\parallel}\right)\Phi(x_\perp) = \rho(s_{e_p} s_x)\ .\label{other-conditions-2}
\end{align}
Let us consider the case of an scalar bulk field and try to put $\mu$ to be the trivial representation. Then the equations \eqref{other-conditions-1}  and \eqref{other-conditions-2}  give only one non-trivial condition
\begin{equation}
    \Phi(r_q x_\perp) =\Phi(x_\perp)\ .
\end{equation}
Combining it with \eqref{first-conditions}  we arrive at the unique solution
\begin{equation}\label{scalar-phi}
    \Phi(x) = |x_\perp|^\Delta\ .
\end{equation}
Thus, we have constructed an isomorphism between scalar fields in the bulk and a class of covariant functions on $G_{d,p}$. As this is the only setup that we will consider in applications of later sections, we will not discuss extensions to the case of spinning bulk fields at present.
\medskip

{\bf Example} Let us illustrate parts of the above discussion on the simplest non-trivial example, that of a line defect in a two-dimensional conformal field theory. The conformal group of the Euclidean plane is $G_d=SO(3,1)$ and the defect group is $G_{d,p} = SO(2,1)$. Let $\mathfrak{g}_d = \mathfrak{so}(3,1)$ be the complexified Lie algebra of $G_d$. We choose its basis
\begin{equation}\label{generators}
    \mathfrak{g}_d = \text{span}\{P_\mu,K_\mu,D,M\},\quad \mu=1,2,
\end{equation}
with the non-zero bracket relations written in the appendix A. The notation here is $M_{12} = -M_{21} = M$. The representation on fields of conformal weight $\Delta$ and spin $l$ by differential operators reads
\begin{equation}\label{2d-operators}
    p_\mu = \partial_\mu,\quad m_{\mu\nu} = x_\nu \partial_\mu - x_\mu \partial_\nu - l\epsilon_{\mu\nu},\quad d = x^\mu \partial_\mu + \Delta,\quad k_\mu = x^2\partial_\mu - 2x_\mu d - 2l x^\nu \epsilon_{\mu\nu}\ .
\end{equation}
We use the summation convention and raise and lower the indices with a flat Euclidean metric. The Levi-Civita symbol has $\epsilon_{12}=1$.  Differential operators satisfy the opposite brackets compared to the generators \eqref{generators}. The defect algebra is spanned by $\{P_1,K_1,D\}$ and is isomorphic to the Lie algebra of the conformal group in one dimension. We write the differential operators explicitly
\begin{equation}\label{1d-operators}
    p_1 = \partial_1,\quad d = x_1\partial_1 + x_2\partial_2 + \Delta, \quad k_1 = (x_2^2 - x_1^2)\partial_1 - 2 x_1 x_2 \partial_2 - 2\Delta x_1 - 2 l x_2\ .
\end{equation}
These operators should be compared with the right-invariant vector fields on $SL(2)$ in the coordinates specified by
\begin{equation}
    g = g_d(x) e^{\mu(P_1-K_1)} =  e^{x_1 P_1} e^{\log x_2 D} e^{\mu(P_1-K_1)}\ .
\end{equation}
The vector fields are computed form the Maurer-Cartan form to give
\begin{equation}\label{2d-example}
    \tilde p = \partial_1, \quad \tilde d = x_1\partial_1 + x_2 \partial_2, \quad \tilde k = (x_2^2 - x_1^2)\partial_1 - 2 x_1 x_2 \partial_2 - x_2\partial_\mu\ .
\end{equation}
If we look at scalar fields, the construction above instructs us to set $\partial_\mu = 0$ and conjugate the operators 
\eqref{2d-example}  by $|x_2|^{-\Delta}$ in order to obtain \eqref{1d-operators}  with $l=0$. A simple calculation verifies 
that this is indeed the case.

\section{Lifting Correlation Functions}

In this section we will use the lifts of bulk and defect fields to write down a new representation of correlation functions as functions on a number of copies of the defect conformal group. A correlator of $m$ bulk and $n$ defect fields will be written in terms of a covariant function on $G_{d,p}^{m+n}$. Our goal is to eventually end up with functions on just one copy of $G_{d,p}$ which can be done if the number of insertion points is sufficiently small. As a first step in this direction, we will show in the second subsection how one can lift pairs of bulk and defect fields, a trick that will be useful when we come to analyse three-point functions later. The third subsection treats the example of two-point functions involving one bulk and one defect field as a simple illustration of the formalism.

\subsection{Ward identities}

In this subsection we study the form of correlation functions in a conformal theory on $\mathbb{R}^d$, in the presence of a $p$-dimensional defect. A correlation function involving $m$ bulk and $n$ defect fields is denoted by
\begin{equation}
    G_{m,n}(x_1,...,x_m,\hat x_1,...,\hat x_n) = \langle \varphi_1(x_1)...\varphi_m(x_m)\hat\varphi_1(\hat x_1)...\hat\varphi_n(\hat x_n)\rangle\ .
\end{equation}
Let $\rho_1,...,\rho_m,\hat\rho_1,...,\hat\rho_n$ be representations that label the fields entering the correlation function. Bulk fields are labelled by representations of the group $K_d$, while the defect fields are labelled by representations of $K_p\times SO(q)$. The Ward identities read
\begin{equation}\label{Ward-identities}
    G_{m,n}(h x_1,...,h x_m, h \hat x_1,...,h \hat x_n) = \Big(\rho_1(k(x_1,h))\otimes...\otimes \hat\rho_n(k(\hat x_n,h))\Big) G_{m,n}(x_1,...,\hat x_n)\ .
\end{equation}
Notice that if $x = \hat x$ lies on the defect, $k(\hat x_i,h)\in K_p\times SO(q)$, so it makes sense to evaluate the representation $\hat \rho_i$ at this element. We can reformulate eq.\ \eqref{Ward-identities}  by saying that $G_{m,n}$ is an invariant vector in the representation $\pi_1\otimes...\otimes\hat\pi_n$ of the defect conformal group
\begin{equation}\label{invariant-vector}
    G_{m,n} = (\pi_1(h)\otimes...\otimes\hat\pi_n(h)) G_{m,n}\ .
\end{equation}
Representations $\pi_i$ were defined in the previous section and the definition of $\hat\pi_j$ is entirely analogous.

Let us now assume that we are given two sets of intertwiners $(g_d,S_{d,p},\mu_i,\Phi_i)$ and $(g_p,\hat S_{d,p},\hat\mu_j,\hat\Phi_j)$ as described in the last section (in particular $\hat\Phi_j=1$). They allow to lift any solution to the Ward identities $G_{m,n}$ to a function $F_{m,n}:G_{d,p}^{m+n}\xrightarrow{}V$ ($V$ is the tensor product of spaces of polarisations of the fields) which satisfies
\begin{equation}\label{Fmn-section}
    F_{m,n}(g_d(x_1),...,g_p(\hat x_n)) = \Phi_1(x_1)...\hat\Phi_n(\hat x_n) G_{m,n}(x_1,...,\hat x_n),
\end{equation}
and is right covariant
\begin{align}\label{right-covariance-Fmn}
    & F_{m,n}(g_1 s_1,...,g_m s_m, g_{m+1} \hat s_{1},...,g_{m+n}\hat s_n) =\\
    & = \Big( \mu_1(s_1^{-1})\otimes...\otimes\mu_m(s_m^{-1})\otimes\hat\mu_1(\hat s_1^{-1})\otimes...\otimes\hat\mu_n(\hat s_n^{-1}) \Big) F_{m,n}(g_1,...,g_{m+n})\ .
\end{align}
These two properties ensure $F_{m,n}$ is defined almost everywhere on $G_{d,p}^{m+n}$. Now, the invariance of $G_{m,n}$, eq.\ \eqref{invariant-vector}, and the intertwining property of lifts imply that $F_{m,n}$ is
invariant under the diagonal left regular action of $G_{d,p}$
\begin{equation}
     F_{m,n}(h g_i) = F_{m,n}(g_i)\ .
\end{equation}
The function $F_{m,n}$ is our new representation of the correlator and the starting point for several other representations that will be constructed below.

\subsection{Pairing up bulk and defect fields}

In considerations of correlation functions $G_{m,n}$ we will find it useful to pass from functions $F_{m,n}$ to functions on a smaller number of copies of the defect conformal group. We can efficiently achieve this by "pairing up bulk and defect points", thus effectively lifting pairs of fields, rather than individual ones. We now explain this process in more detail.

As we have mentioned, functions $f, \hat f$ belong to induced representations of the defect conformal group
\begin{equation}
    f\in \pi =\text{Ind}_{S_{d,p}}^{G_{d,p}} W, \quad \hat f\in\hat \pi =\text{Ind}_{\hat S_{d,p}}^{G_{d,p}} \hat W\ .
\end{equation}
We have used the same notation $\pi,\hat\pi$ as for the representations on fields because our analysis indeed showed that these representations are isomorphic to one another. The tensor product $\pi\otimes\hat\pi$ is naturally realised in the space of functions
\begin{equation}
    F : G_{d,p}^2 \xrightarrow{} W\otimes\hat W, \quad F(g_1 s, g_2 \hat s) = \left(\mu(s)^{-1}\otimes\hat\mu(\hat s)^{-1}\right) F(g_1,g_2),
\end{equation}
under the diagonal left-regular action. We will be interested in another way of realising this representation:
\smallskip

\noindent
{\bf Proposition} Let $K=SO(p)\times SO(q-1)$ be the stabiliser of a pair of one bulk and one defect point in $G_{d,p}$. The following is an isomorphism of $G_{d,p}$-modules
\begin{align}
     Q : \pi\otimes\hat\pi\xrightarrow{}\chi=\text{Ind}_{K}^{G_{d,p}}(\mu\otimes\hat\mu),\quad (Q F)(g) = F(g,g)\ .
\end{align}
Thus, $Q$ is essentially composing a function $F$ with the coproduct map on the group algebra $L^1(G)$.\footnote{We will not be precise about the classes of functions on which groups act in this paper. The interested reader is referred to \cite{Kirillov}.} It is the properties of the coproduct that ensure $Q$ respects the $G_{d,p}$-action.
\smallskip

{\it Proof}: First, observe that the representation $\chi$ is well-defined. Indeed, both $\mu$ and $\hat\mu$ are representations of $K$ by restriction. Let us show that $Q(F)\in\chi$, that is, that is has the required covariant properties
\begin{equation}
    Q(F)(g k) = F(g k, g k) = \left(\mu(k)^{-1}\otimes\hat\mu(k)^{-1}\right) F(g,g) = (\mu\otimes\hat\mu)(k^{-1}) Q(F)(g)\ .
\end{equation}
Thus, $Q$ is well-defined. It is clearly a $G_{d,p}$-module homomorphism
\begin{equation}
    Q(g\cdot F)(g') = (g\cdot F)(g',g') = F(g g',g g') = Q(F)(gg') = (g\cdot Q(F))(g')\ .
\end{equation}
It remains to prove that $Q$ is a bijection. To this end, notice that almost any element $g\in G_{d,p}$ can be written as $g = \hat s s$ with $\hat s\in \hat S_{d,p}$ and $s\in S_{d,p}$. This is true by the following argument
\begin{equation}
    g = g_p g_q = n_I(g_p) a_I(g_p) k_I(g_p) g_q = n_I(g_p) a_I(g_p) g_q\ k_I(g_p)\ .
\end{equation}
In the first step, we have written $g$ as a product of elements in $SO(p+1,1)$ and $SO(q)$. Then we have factorised the first term according to the Iwasawa decomposition and moved $g_q$, past $k_I(g_p)$. The last expression is of the correct form $\hat s s$.

We can now reconstruct $F$ from $Q(F)$. Given two elements $g_1, g_2\in G_{d,p}$, let $s_1,\hat s_2,$ be the above solutions to the decomposition $g_2^{-1} g_1  = \hat s_2 s_1^{-1}$. Then we have
\begin{align*}
     F(g_1,g_2) = \left(\mu(s_1)\otimes\hat\mu(\hat s_2)\right) F(g_1 s_1,g_2 \hat s_2) = \left(\mu(s_1)\otimes\hat\mu(\hat s_2)\right) F(g_1 s_1,g_1 s_1) = \left(\mu(s_1)\otimes\hat\mu_2(\hat s_2)\right) Q(F)(g_1 s_1).
\end{align*}
This completes the proof of the proposition.
\smallskip 

As a consequence, we can lift a pair of primary fields, one bulk and one defect, by composing the individual lifts with 
the isomorphism $Q$. The constructions allows us to uplift correlation functions of $m$ bulk and $n$ defect fields to a 
product group with just $\textit{max}(m,n)-1$ factors $G_{d,p}$. In particular, correlation functions of $m=2$ 
bulk and $n=2$ defect fields can be uplifted to functions on a single copy of the defect conformal group. 

\subsection{An example: bulk-defect two-point function}

As a simple example of the above ideas, let us determine the form of a two-point function of one bulk and one defect 
field, $G_{1,1}(x_i)$. The function $G_{1,1}(x_i)$ lifts to a function $F_{1,1}:G_{d,p}^2\xrightarrow{}V$ which satisfies
\begin{equation}
    F_{1,1}(g_1 s_1,g_2\hat s_2) = \left( \mu_1(s_1^{-1})\otimes\hat\mu_2(\hat s_2^{-1}) \right) F_{1,1}(g_1,g_2), \quad F_{1,1} (h g_i) = F_{1,1}(g_i)\ .
\end{equation}
Let us put $F = Q(F_{1,1})$ , that is $F(g) = F_{1,1}(g,g)$. Then $F$ is a constant function
\begin{equation}
    F(h g) = F_{1,1}(h g,h g) = F_{1,1}(g,g) = F(g)\ .
\end{equation}
To write the two-point function in terms of $F$ we need to Iwasawa-decompose
\begin{equation}
    g_{12}(x_i) = g_p(\hat x_2)^{-1} g_d (x_1) = e^{(x_1^a - \hat x_2^a)P_a} |x_{1\perp}|^D e^{\varphi^i M_{id}} \equiv \hat s_{12} s_{12}^{-1}\ .
\end{equation}
As explained in the previous section, $s_{12}$ and $\hat s_{12}$ are essentially the Iwasawa factors of $g_{12}$. We have
\begin{align*}
    & e^{\hat x_{12}^a P_a} |x_{1\perp}|^D = |x_{1\perp}|^D e^{\frac{\hat x_{12}^a}{|x_{1\perp}|}P_a} = |x_{1\perp}|^D e^{\frac{|x_{1\perp}|\hat x_{12}^a}{x_{1\perp}^2+\hat x_{12}^2}K_a}\left(\frac{\hat x_{12}^2+x_{1\perp}^2}{x_{1\perp}^2}\right)^{D} k_I\left(-\frac{\hat x_{12}}{|x_{1\perp}|}\right)\\
    & = e^{\frac{\hat x_{12}^a}{x_{1\perp}^2+\hat x_{12}^2}K_a}\left(\frac{\hat x_{12}^2+x_{1\perp}^2}{|x_{1\perp}|}\right)^D k_I\left(\frac{\hat x_{21}}{|x_{1\perp}|}\right)\ .
\end{align*}
Therefore, the factors are
\begin{equation}\label{s-and-shat}
    \hat s_{12} = e^{\frac{\hat x_{12}^a}{x_{1\perp}^2+\hat x_{12}^2}K_a}\left(\frac{\hat x_{12}^2+x_{1\perp}^2}{|x_{1\perp}|}\right)^D e^{\varphi_1^i M_{id}}, \quad s_{12} = k_I\left(\frac{\hat x_{21}}{|x_{1\perp}|}\right)^{-1}\ .
\end{equation}
In terms of $F$, the two point function reads
\begin{equation}
    G_{1,1}(x_i) = \frac{1}{\Phi_1(x_1)} F_{1,1}(g_d(x_1), g_p(\hat x_2)) = \frac{1}{\Phi_1(x_1)}\left(\mu_1(s_{12})\otimes\hat\rho_2(\hat s_{12})\right) F\left(g_d(x_1) s_{12}\right)\ .
\end{equation}
Let us evaluate this expression further in the case where the bulk field is a scalar. Then we should put $\mu_1=1$, hence
\begin{equation}
    G_{1,1}(x_i) = c\frac{(x_{1\perp}^2+\hat x_{12}^2)^{-\Delta_{\hat 2}}}{|x_{1\perp}|^{\Delta_{1\hat 2}}} \hat\rho_2(e^{\varphi_1^i M_{id}}),
\end{equation}
for some constant $c$ (such that $F\equiv c$). We have written the conformal dimensions of two fields as $\Delta_1$ and $\Delta_{\hat 2}$ and used the shorthand notation $\Delta_{1\hat2} = \Delta_1 - \Delta_{\hat 2}$. If one assumes the transverse (internal) spin of the second field to be trivial, i.e.\ $\hat\rho_2(e^{\varphi_1^i M_{id}})= 1$, one recognises the usual expression for the two-point function.

\section{Conformal Blocks from Harmonic Analysis}

This section is dedicated to the study of conformal blocks. Since the influential work of Dolan and Osborn, these are usually characterised as eigenfunctions of appropriate Casimir operators. Under a lift, the action of the defect conformal group on the bulk space or the defect is carried to the left regular action. Thus, the quadratic Casimir is carried to the Laplace-Beltrami operator on $G_{d,p}$.

In the previous section, we explained how to lift arbitrary correlation functions $G_{m,n}$, involving $m$ bulk and $n$ defect fields, to covariant functions $F_{m,n}$ on $m+n$ copies of the group $G_{d,p}$. We have also seen how to lift pairs of bulk and boundary fields. These results allow to map a number of correlation functions with small values of $m$ and $n$ to functions on just one copy of the group $G_{d,p}$ by applying some further simple transformations. As a general rule, one of the Dolan-Osborn-like quadratic Casimirs is carried to the Laplacian under these transformations. As we will see in the first two subsections, for two-point functions of bulk fields as well as three-point functions of two defect and one bulk field, the eigenproblem of this operator together with appropriate boundary conditions completely characterise the conformal blocks. The latter can be constructed in terms of Gauss' hypergeometric function $_2F_1$ and its special incarnation as a Gegenbauer polynomial. In the third subsection, we consider the three-point function of two bulk and one defect field. This leads us to the eigenvalue problem of two coupled second order operators. Upon closer inspection, the system is equivalent to the Appell system so that the conformal blocks can be expressed
in terms of the Appell function $F_4$ (and Gegenbauer polynomials). Our results extend those of \cite{Lauria:2020emq} and reduce to the 
findings of Lauria et al. upon restriction to a 2-dimensional subspace of cross ratios.  

\subsection{Bulk-bulk two-point function}

Let us move to two-point functions of bulk fields. Their kinematical form is no longer completely fixed by symmetry and there are two invariants on which they can depend. According to the general theory, the two-point function $G_{2,0}(x_i)$ lifts to a function $F_{2,0}:G_{d,p}^2\xrightarrow{}V$ which satisfies
\begin{equation}
    F_{2,0}(g_1 s_1,g_2 s_2) = \left( \mu_1(s_1^{-1})\otimes\mu_2( s_2^{-1}) \right) F_{2,0}(g_1,g_2), \quad F_{2,0} (h g_i) = F_{2,0}(g_i)\ .
\end{equation}
Let us define a function $F:G_{d,p}\xrightarrow{}V$ by $F(g) = F_{2,0}(e,g)$. We can easily recover $F_{2,0}$ from $F$ using the above covariance properties
\begin{equation*}
    F_{2,0}(g_1,g_2) = F_{2,0}(e,g_1^{-1}g_2) = F(g_1^{-1} g_2)\ .
\end{equation*}
On the other hand, $F$ is left-right covariant with respect to the subgroup $S_{d,p}\subset G_{d,p}$
\begin{equation}
    F(s_1 g s_2) = F_{2,0}(e,s_1 g s_2) = F_{2,0}(s_1^{-1},g s_2) = \left( \mu_1(s_1)\otimes \mu_2(s_2^{-1})\right) F_{2,0}(e,g) = \left( \mu_1(s_1)\otimes \mu_2(s_2^{-1})\right) F(g)\ .\nonumber
\end{equation}
Therefore $F$ can be regarded as a function on the double quotient $M_{2pt} = S_{d,p}\backslash G_{d,p}/S_{d,p}$. This space is two-dimensional as almost any element of $G_{d,p}$ can be written in the form
\begin{equation}\label{2pt-decomposition}
    g = r^{p+1}_l e^{\lambda D} r^{p+1}_r\ r^{q-1}_l e^{\kappa M_{d-1,d}} r^{q-1}_r\ ,
\end{equation}
with $r^{p+1}_{l,r}\in SO(p+1)$ and $r^{q-1}_{l,r}\in SO(q-1)$. We will refer to this factorisation as the a Cartan decomposition of $g$. The space $M_{2pt}$ is the direct product of two double quotients of similar forms, $SO(p+1)\backslash SO(p+1,1)/SO(p+1)$ and $SO(q-1)\backslash SO(q)/SO(q-1)$, each of which is one-dimensional. The Cartan decomposition is of course not unique, but this fact has no bearing on our discussion. The function $F$ satisfies
\begin{equation}\label{2pt-covariance}
    F(g) = \left(\mu_1(r^{p+1}_l r^{q-1}_l)\otimes\mu_2(r^{p+1}_r r^{q-1}_r)^{-1}\right) F(e^{\lambda D + \kappa M_{d-1,d}})\ .
\end{equation}
The restriction of $F$ to the two-dimensional abelian subgroup generated by $D$ and $M_{d-1,d}$ will be denoted by $\psi(\lambda,\kappa) = F(e^{\lambda D + \kappa M_{d-1,d}})$. We can relate $\psi$ and the two point function as soon as the Cartan decomposition of $g_d(x_1)^{-1} g_d(x_2)$ is known
\begin{equation}
    G_{2,0}(x_i) = \frac{1}{\Phi_1(x_1)\Phi_2(x_2)} F_{2,0}(g_d(x_1), g_d(x_2)) = \frac{1}{\Phi_1(x_1)\Phi_2(x_2)} F\left(g_d(x_1)^{-1} g_d(x_2)\right)\ .
\end{equation}
Let us denote the argument of $F$ by $g_{2pt}(x_i)$. We determine its Cartan factors
\begin{align}\label{lambda-coordinate}
     g_{2pt}(x_i)_p = |x_{1\perp}|^{-D} e^{x_{21}^a P_a} |x_{2\perp}|^D = r^{p+1}_l(x_i) e^{\lambda D} r^{p+1}_r(x_i) \quad\text{with}\quad \cosh\lambda = \frac{x_{1\perp}^2 + x_{2\perp}^2 +\hat x_{12}^2}{2|x_{1\perp}| |x_{2\perp}|}\ .
\end{align}
For a simple proof of eq. \eqref{lambda-coordinate} , see the appendix B. The factors $r^{p+1}_{l,r}$ are computed similarly. Let us now turn to the $SO(q)$-part. Again, in the appendix B, it is shown that
\begin{align}\label{kappa-coordinate}
    g_{2pt}(x_i)_q = e^{-\varphi_1^i M_{id}} e^{\varphi_2^j M_{jd}} = r_l^{q-1}(x_i) e^{\kappa M_{d-1,d}} r_r^{q-1}(x_i) \quad \text{with} \quad \cos\kappa = \frac{x_1^i x_2^i}{|x_{1\perp}| |x_{2\perp}|}\ .
\end{align}
Hence, the correlation function becomes
\begin{equation}
    G_{2,0}(x_i) = \frac{1}{\Phi_1(x_1)\Phi_2(x_2)}\left(\mu_1(r^{p+1}_l(x_i) r^{q-1}_l(x_i))\otimes\mu_2(r^{p+1}_r(x_i) r^{q-1}_r(x_i))^{-1}\right) \psi(\lambda,\kappa)\ .
\end{equation}
Let us evaluate this expression further in the case when the fields are scalar. Then we should put $\mu_i=1$, so
\begin{equation}\label{2pt-bulk-scalar}
    G_{2,0}(x_i) = \frac{1}{|x_{1\perp}|^{\Delta_1} |x_{2\perp}|^{\Delta_2}}\psi(\lambda,\kappa)\ .
\end{equation}
The coordinates $(\lambda,\kappa)$ are the two independent conformal invariants. They are related to coordinates $(\phi,\chi)$ used in \cite{Billo:2016cpy} by
\begin{equation}\label{2pt-map-to-Lauria}
    \kappa = \phi,\quad \cosh\lambda = \frac12 \chi\ .
\end{equation}
We recognise in eq.\ \eqref{2pt-bulk-scalar}  the usual expression for the two point function.

Conformal blocks are eigenfunctions of the Laplace-Beltrami operator within the space of covariant functions \eqref{2pt-covariance} . Let us show how this comes about. Partial waves for the two-point function $G_{2,0}(x_i)$ can be characterised as eigenfunctions of the quadratic Casimir that is constructed out of the vector fields that represent the action of the defect conformal algebra $\mathfrak{g}_{d,p}$ on a scalar field. We may chose either the first or the second point for these differential operators. Let us choose the second one to be concrete. After the correlation function is lifted to $F_{2,0}$, results of the previous section tell us that the action generated by these vector fields maps to the left-regular action {\it on the second copy} of $G_{d,p}$. The corresponding geometric representation on $F_{2,0}$ reads
\begin{equation}
    (g\cdot F_{2,0})(g_1,g_2) = F_{2,0}(g_1,g^{-1}g_2)\ .
\end{equation}
This is indeed a representation of $G_{d,p}$ on the space $L^1(G_{d,p}^2,V)$, but clearly it does not respect the covariance properties satisfied by $F_{2,0}$. However, the quadratic Casimir does respect the covariance properties and equals the Riemannian Laplace-Beltrami operator $\Delta$ on the second copy of $G_{d,p}$, denoted $\Delta^{(2)}$. Furthermore, by the definition of $F$
\begin{equation}
    \Delta F = \Delta^{(2)} F_{2,0},
\end{equation}
and hence the Casimir operator acting on $F$ coincides with the Laplacian, as claimed. Conformal blocks factorise according to the direct product structure of $G_{d,p}$. It is possible to write the restriction $\psi_{\hat\Delta,s}(\lambda,\kappa) = \psi_{p,\hat\Delta}(\lambda)\psi_{q,s}(\kappa)$ using standard representation theory as in \cite{Vilenkin}. For scalar fields, conditions \eqref{2pt-covariance}  tell us that $\psi_{p,\hat\Delta},\psi_{q,s}$ are zonal spherical functions
\begin{equation}\label{2pt-blocks}
    \psi_{p,\hat\Delta} = (\cosh\lambda)^{-\hat\Delta}\ _2 F_1\left(\frac{\hat\Delta+1}{2},\frac{\hat\Delta}{2};\frac{p+1}{2};\tanh^2\lambda\right), \quad \psi_{q,s}(\kappa) = \frac{s! (q-3)!}{(s+q-3)!} C^{(q-2)/2}_s (\cos\kappa)\ .
\end{equation}
Here $C^{(q-2)/2}_s$ is the Gegenbauer polynomial. The function $\psi_{p,\hat\Delta}$ can be expressed in terms of a Legendre function using a hypergeometric identity. In fact, the functions $\psi_{p,\hat\Delta}$ and $\psi_{q,s}$ are very similar to each other, which is clear from the fact that they come from quotients that are related by analytic continuation. See \cite{Vilenkin} for more details.

Let us compare our conformal blocks to those of \cite{Billo:2016cpy}. For the transverse part, we observe that the polynomials $\psi_{q,s}(\kappa)$ readily agree with the functions $(4.9)$ from this paper. As for $\psi_{p,\hat\Delta}$, notice that the function
\begin{equation}
    \tilde \psi_{p,\hat\Delta} = (\cosh\lambda)^{-\hat\Delta}\ _2 F_1\left(\frac{\hat\Delta+1}{2},\frac{\hat\Delta}{2};1+\hat\Delta-\frac{p}{2};1-\tanh^2\lambda\right),
\end{equation}
solves the same hypergeometric equation as $\psi_{p,\hat\Delta}$. Indeed, in our discussion above, we did not include the analysis of boundary conditions that supplement the Casimir differential equation. Once this is done, it turns out the $\tilde\psi_{p,\hat\Delta}$ is the correct eigenfunction to use. Now from eq.\ \eqref{2pt-map-to-Lauria} we can rewrite
\begin{equation}
    \tilde \psi_{p,\hat\Delta} = \left(\frac{\chi}{2}\right)^{-\hat\Delta}\ _2 F_1\left(\frac{\hat\Delta+1}{2},\frac{\hat\Delta}{2},1+\hat\Delta-\frac{p}{2},\frac{4}{\chi^2}\right),
\end{equation}
in agreement with eq.\ $(4.7)$ of \cite{Billo:2016cpy}. This concludes our analysis of two-point correlation 
functions of scalar bulk fields in the presence of a defect. Of course our results here are not new, but the 
well studied setup can illustrate nicely how the group theoretic approach works.   

\subsection{Bulk-defect-defect three-point function}

In order to address the less studied example of a three-point function involving one bulk and two defect 
fields, we start with our familiar lift of the correlator $G_{1,2}(x_i)$ to a function $F_{1,2}:G_{d,p}^3
\xrightarrow{}V$ which satisfies
\begin{equation}
    F_{1,2}(g_1 s_1,g_2\hat s_2,g_3 \hat s_3) = \left( \mu_1(s_1^{-1})\otimes\hat\mu_2(\hat s_2^{-1})
    \otimes\hat\mu_3(\hat s_3^{-1}) \right) F_{1,2}(g_1,g_2,g_3), \quad F_{1,2} (h g_i) = F_{1,2}(g_i)\ .
\end{equation}
To simplify the lift, we can pair up the two defect fields in the correlation function and set $F(g) = 
F_{1,2}(e,g,g w_p^{-1})$. The function $F_{1,2}$ is now reconstructed from $F$ as
\begin{equation}
    F_{1,2}(g_1,g_2,g_3) = \left(\hat\mu_2(\hat s_2)\otimes\hat\mu_3(\hat s_3)\right) 
    F_{1,2}(g_1, g_2\hat s_2, g_3 \hat s_3) = \left(\hat\mu_2(\hat s_2)\otimes\hat\mu_3(\hat s_3)
    \right) F(g_1^{-1} g_2 \hat s_2),
\end{equation}
where the elements $\hat s_2, \hat s_3\in \hat S_{d,p}$ solve the equation $g_3 \hat s_3 = g_2 \hat 
s_2 w_p^{-1}$. The function $F$ has left-right covariance properties
\begin{equation*}
    F(s g l) = F_{1,2}(e,s g l,s g l w_p^{-1}) = F_{1,2}(s^{-1},g l,g w_p^{-1} w_p l w_p^{-1}) = \left(\mu_1(s)\otimes\hat\mu_2(l^{-1})\otimes\hat\mu_3(w_p l^{-1} w_p^{-1})\right) F(g),
\end{equation*}
with $l\in L = SO(1,1)\times SO(p)\times SO(q)$. The way we constructed $F$ mimics the pairing up 
of points in a bulk conformal field theory without defect insertion that was done in 
\cite{Schomerus:2016epl}. Indeed, there one associates to a pair of fields in $\mathbb{R}^d$ a 
$K_d$-covariant function on the conformal group. The group $K_d$ naturally appears as the stabiliser 
of the pair $(0,\infty)$. Here the same construction is performed along the defect, while transverse 
directions play no role. The function $F$ can be regarded as a function on the coset space $S_{d,p}
\backslash G_{d,p}/ L$. This space is one-dimensional
\begin{equation}
    X = S_{d,p} \backslash G_{d,p}/ L = SO(p+1)\backslash SO(p+1,1) / \left(SO(1,1)\times SO(p)\right)\ .
\end{equation}
We parametrise it by writing group elements as
\begin{equation}
    g = k_I e^{y K_1} r^p e^{\lambda D}\ r^q,
\end{equation}
with $k_I\in SO(p+1),\ r^p\in SO(p)$ and $r^q\in SO(q)$. The function $F$ is determined by its restriction to the group generated by $K_1$
\begin{equation}
    F(g) = \left( \mu_1(k_I) \otimes \hat\mu_2 (r^p e^{\lambda D} r^q)^{-1} \otimes \hat\mu_3 (w_p r^p e^{\lambda D} r^q w_p^{-1})^{-1}\right) \varphi(y)\ .
\end{equation}
We have denoted the restriction by $\varphi(y)=F(e^{yK_1})$. The correlation function is related to $F$ by
\begin{equation}\label{corr}
    G_{1,2}(x_i) = \frac{1}{\Phi_1(x_1)} F_{1,2}(g_d(x_1),g_p(\hat x_2),g_p(\hat x_3)) = \frac{1}{\Phi_1(x_1)} \left(\hat\mu_2(\hat s_2)\otimes\hat\mu_3(\hat s_3)\right) F(g_d(x_1)^{-1}g_p(\hat x_2) \hat s_2)\ .
\end{equation}
Group elements $\hat s_2$ and $\hat s_3$ are determined in the non-defect theory on $SO(p+1,1)$, \cite{Buric:2020buk}, and read
\begin{equation}
    \hat s_2 = w_p^{-1} m(w_p\hat x_{32}) w_p, \quad \hat s_3 = k_p(t_{32})^{-1} w_p^{-1} m(\hat x_{32}) w_p\ .
\end{equation}
Our notation here coincides with the one used in that paper. In particular $t_{32}$ stands for $t(\hat x_{32})$. Therefore, the argument of $F$ has the $SO(p+1,1)$-part
\begin{align*}
    (g_d(x_1)^{-1}g_p(\hat x_2) \hat s_2)_p & = |x_{1\perp}|^{-D} e^{\hat x_{21}^a P_a} e^{\frac{\hat x_{32}^a}{\hat x_{32}^2} K_a} = e^{\frac{\hat x_{21}^a}{|x_{1\perp}|} P_a} e^{\frac{|x_{1\perp}|\hat x_{32}^a}{\hat x_{32}^2} K_a} |x_{1\perp}|^{-D}\\
    & = k_I\left(\frac{\hat x_{12}}{|x_{1\perp}|}\right) \left(\frac{\hat x_{12}^2 + x_{1\perp}^2}{x_{1\perp}^2}\right)^{-D} e^{\left(\frac{|x_{1\perp}|\hat x_{21}^a}{\hat x_{12}^2 + x_{1\perp}^2}+\frac{|x_{1\perp}|\hat x_{32}^a}{\hat x_{32}^2}\right)K_a}|x_{1\perp}|^{-D}\\
    & = k_I\left(\frac{\hat x_{12}}{|x_{1\perp}|}\right) e^{\frac{1}{|x_{1\perp}|}\left(\hat x_{21}+(\hat x_{12}^2 + x_{1\perp}^2)\frac{\hat x_{32}}{\hat x_{32}^2}\right)^a K_a}\left(\frac{\hat x_{12}^2 + x_{1\perp}^2}{|x_{1\perp}|}\right)^{-D}\ .
\end{align*}
In the first step we used the result \eqref{Iwasawa-2}. The coordinate $y$ may be read off as 
\begin{equation}
    y = \frac{1}{|x_{1\perp}| |\hat x_{23}|} \sqrt{(\hat x_{12}^2 + x_{1\perp}^2)(\hat x_{13}^2 + x_{1\perp}^2) - x_{1\perp}^2 \hat x_{23}^2}  = \sqrt{u_{23,1}^{-1} - 1},
\end{equation}
where the cross ratio $u_{23,1}$ is defined as in \cite{Lauria:2017wav}\footnote{Cross ratios $u_{i,jk},\ u_{ij,k}$ used in this work may differ from those of \cite{Lauria:2017wav} by factors such as $2,-1$ etc. In all formulas, these functions mean the ones explicitly defined in the present paper.}
\begin{equation}
    u_{23,1} = \frac{x_{1\perp}^2 \hat x_{23}^2}{(\hat x_{12}^2+x_{1\perp}^2)(\hat x_{13}^2 + x_{1\perp}^2)}\ .
\end{equation}
For $SO(p)$-scalar fields, the correlation function \eqref{corr}  is further evaluated
\begin{equation}\label{bdd-scalar-2nd-channel}
    G_{1,2}(x_i) = \frac{1}{|x_{1\perp}|^{\Delta_1}} \frac{1}{\hat x_{23}^{2\Delta_{\hat 3}}} \left(\frac{\hat x_{12}^2 + x_{1\perp}^2}{|x_{1\perp}|}\right)^{\Delta_{\hat 3\hat 2}}(\hat\rho_2\otimes\hat\rho_3)(e^{\varphi_1^i M_{id}}) \varphi(y)\ .
\end{equation}
Let us now solve for eigenfunctions of the Laplacian on the space of left-right covariant functions $F$. From the simple relation
\begin{equation}
    \Delta^{(1)} F_{1,2} = \Delta^{(23)} F_{1,2} \mapsto \Delta F,
\end{equation}
under the above mapping, it follows that these eigenfunctions are conformal blocks. Here, $\Delta^{(23)}$ is the quadratic Casimir constructed from the vector fields that generate the diagonal left-regular action on the last two copies of $G_{d,p}$.  We can first consider only the left quotient $S_{d,p}\backslash G_{d,p}$ and parametrise it according to
\begin{equation}
    g = k_I e^{y^a K_a} e^{\lambda D}\ .
\end{equation}
Since the $SO(q)$-factor will be trivialised by the right quotient we omitted writing it in the above equation. Later we will trade $(y^a)$ for spherical polar coordinates $(y,\phi)$ on $\mathbb{R}^p$. We compute the Laplacian from the left-invariant vector fields - ones that generate the right regular action. The action of dilations, special conformal transformations and rotations is simple
\begin{align}
    & k_I e^{y^a K_a} e^{\lambda D} e^{\mu D} = k_I e^{y^a K_a} e^{(\lambda+\mu) D}, \quad k_I e^{y^a K_a} e^{\lambda D} e^{z^a K_a} = k_I e^{(y^a + e^{-\lambda} z^a) K_a} e^{\lambda D},\\
    & k_I e^{y^a K_a} e^{\lambda D} r^p = k_I r^p e^{((r^p)^{-1}y)^a K_a} e^{\lambda D}
\end{align}
Finally, the action of translations is found by the following calculation
\begin{align*}
    & k_I e^{y^a K_a} e^{\lambda D} e^{z^b P_b} = k_I w_p e^{(s_p y)\cdot P} w_p e^{e^{\lambda}z\cdot P} e^{\lambda D}\\
    &= k_I' e^{s_p(y + z^{-2}e^{-\lambda}z)\cdot P} e^{-e^{-\lambda} s_p z\cdot K} s_p s_z\left(\frac{1}{e^{2\lambda}z^2}\right)^D e^{\lambda D} = k_I'' e^{(s_z y - z^{-2}e^{-\lambda}z)\cdot P} e^{e^\lambda z\cdot K} \left(\frac{1}{e^{\lambda}z^2}\right)^D\\
    & = k_I''' e^{(s_z y - z^{-2}e^{-\lambda}z + (1+(s_z y - z^{-2} e^{-\lambda} z)^2)e^\lambda z)\cdot K}\left(\frac{1}{(1+(s_z y - z^{-2} e^{-\lambda} z)^2)e^{\lambda}z^2}\right)^D\ .
\end{align*}
Group elements $k_I',k_I'',k_I'''$ all belong to $SO(p+1)$ and their precise does not matter for the action on the coset. By linearising the above action, the Lie algebra is found to be represented by differential operators
\begin{equation}
    k_a = e^{-\lambda}\partial_{y^a}, \quad d = \partial_\lambda, \quad m_{ab} = y_a\partial_{y^b} - y_b \partial_{y^a}, \quad p_a = e^\lambda \left((1 + y^2)\partial_{y^a} - 2 y_a \partial_\lambda\right)\ .
\end{equation}
The quadratic Casimir, restricted to functions of $(\lambda,y)$ is computed
\begin{equation}\label{Casimir-y-lambda}
    C_2 = \partial_\lambda^2 + (1+y^2)\partial_y^2 - 2 y \partial_y\partial_\lambda + \left((p+1)y+\frac{p-1}{y}\right)\partial_y - p\partial_\lambda\ .
\end{equation}
To pass to the final quotient, we set $\partial_\lambda\xrightarrow{}\Delta_{\hat2\hat3}$. Therefore, conformal blocks satisfy the eigenvalue equation
\begin{equation}\label{Casimir-eqns-bdd}
    \left((1+y^2)\partial_y^2 + \left((p+1-2\Delta_{\hat2\hat3})y + \frac{p-1}{y}\right)\partial_y + \Delta_{\hat2\hat3}(\Delta_{\hat2\hat3}-p)\right)\varphi = \hat\Delta(\hat\Delta-p)\varphi\ .
\end{equation}
This equation is solved by hypergeometric functions
\begin{equation}\label{bbd-blocks}
    \varphi = A \, _2F_1\left(\frac{p-\hat\Delta-\Delta_{\hat2\hat3}}{2},\frac{\hat\Delta-\Delta_{\hat2\hat3}}{2};\frac{p}{2};-y^2\right)+ B y^{2-p} \, _2F_1\left(\frac{2-\hat\Delta-\Delta_{\hat2\hat3}}{2},\frac{2-p+\hat\Delta-\Delta_{\hat2\hat3}}{2};2-\frac{p}{2};-y^2\right).
\end{equation}
Let us compare our results with conformal blocks from \cite{Lauria:2020emq}. The authors there take the limit $\hat x_3\xrightarrow{}\infty$, in which their coordinate $\hat\chi$ is related to $y$ as $\hat\chi = y^2$. They consider the three-point function
\begin{equation}
    \hat x_3^{2\Delta_{\hat 3}}\langle \mathcal{O}_1(x_1) \mathcal{\hat O}_2(\hat x_2)\mathcal{\hat O}_3(\hat x_3) \rangle \sim \frac{e^{i(s_{\hat2}+s_{\hat3})\varphi_1}}{|x_{1\perp}|^{\Delta_1 + \Delta_{\hat2\hat3}}} \sum\mathcal{F}^{\mathcal{\hat O}_2\mathcal{\hat O}_3}_{\mathfrak{p},s}(\hat\chi)\ .
\end{equation}
Conformal blocks $\mathcal{F}^{\mathcal{\hat O}_2\mathcal{\hat O}_3}_{\mathfrak{p},s}$ read
\begin{equation}
    \mathcal{F}^{\mathcal{\hat O}_2\mathcal{\hat O}_3}_{\mathfrak{p},s}(\hat\chi) = \hat\chi^{-\frac12(\hat\Delta+\Delta_{\hat2\hat3})}\ _2F_1\left(\frac{\hat\Delta+\Delta_{\hat2\hat3}}{2},\frac{2-p+\hat\Delta+\Delta_{\hat2\hat3}}{2},1-\frac{p}{2}+\hat\Delta;-\frac{1}{\hat\chi}\right)\ .
\end{equation}
By expanding around zero instead of infinity, we can rewrite these as
\begin{equation}
    \mathcal{F}^{\mathcal{\hat O}_2\mathcal{\hat O}_3}_{\mathfrak{p},s}(\hat\chi) =\  _2F_1\left(\frac{p-\hat\Delta+\Delta_{\hat2\hat3}}{2},\frac{\hat\Delta+\Delta_{\hat2\hat3}}{2},\frac{p}{2};-\hat\chi\right)\ .
\end{equation}
On the other hand, our expression for $G_{1,2}(x_i)$ becomes in the $\hat x_3\xrightarrow{}\infty$ limit
\begin{equation}
    \hat x_3^{2\Delta_{\hat 3}}G_{1,2}(x_i) = \frac{(\hat\rho_2\otimes\hat\rho_3)(e^{\varphi_1^i M_{id}})}{|x_{1\perp}|^{\Delta_1 + \Delta_{\hat2\hat3}}} (1 + y^2)^{\Delta_{\hat3\hat2}} \varphi(y)\ .
\end{equation}
Therefore, we should have the relation $\mathcal{F}^{\mathcal{\hat O}_2\mathcal{\hat O}_3}_{\mathfrak{p},s} = (1 + y^2)^{\Delta_{\hat3\hat2}} \varphi$. There is a number of ways to verify that this is true. Perhaps the simplest one is to conjugate the operator on the left hand side of eq.\ \eqref{Casimir-eqns-bdd}  by $(1+y^2)^{\Delta_{\hat3\hat2}}$. Then eigenfunctions of this new operator should coincide with $\mathcal{F}^{\mathcal{\hat O}_2\mathcal{\hat O}_3}_{\mathfrak{p},s}$. Indeed, the eigenfunctions read
\begin{equation*}
    A \, _2F_1\left(\frac{p-\hat\Delta+\Delta_{\hat2\hat3}}{2},\frac{\hat\Delta+\Delta_{\hat2\hat3}}{2};\frac{p}{2};-y^2\right)+ B y^{2-p} \, _2F_1\left(\frac{2-\hat\Delta+\Delta_{\hat2\hat3}}{2},\frac{2-p+\hat\Delta+\Delta_{\hat2\hat3}}{2};2-\frac{p}{2};-y^2\right),
\end{equation*}
so putting $A=1$ and $B=0$ gives us $\mathcal{F}^{\mathcal{\hat O}_2\mathcal{\hat O}_3}_{\mathfrak{p},s}$. With this we end the discussion of three-point functions of two defect and one bulk field. 

\subsection{Bulk-bulk-defect three point function}

We move to the three-point function that involves two fields in the bulk and one on the defect. Following the familiar strategy, 
we start by lifting the three point function of two bulk and a defect field, $G_{2,1}(x_i)$ to a function $F_{2,1}:G_{d,p}^3
\xrightarrow{}V$ which satisfies
\begin{equation}
    F_{2,1}(g_1 s_1,g_2 s_2,g_3 \hat s_3) = \left( \mu_1(s_1^{-1})\otimes\mu_2(s_2^{-1})\otimes\hat\mu_3(\hat s_3^{-1}) \right) F_{2,1}(g_1,g_2,g_3), \quad F_{2,1}(h g_i) = F_{2,1}(g_i)\ .
\end{equation}
Let us pair up the last two fields by setting $F'(g_1,g_2) = F_{2,1}(g_1,g_2,g_2)$. Then $F'$ obeys $F'(h g_i) = F'(g_i)$ and we put $F(g) = F'(e,g) = F_{2,1}(e,g,g)$. In particular, this implies solutions to $\Delta F = c F$ will correspond to eigenfunctions of the quadratic Casimir acting at the point $x_1$. One reconstructs $F_{2,1}$ from $F$ by
\begin{equation}
    F_{2,1}(g_1,g_2,g_3) =\left(\mu_2(s_2)\otimes\hat\mu_3(\hat s_3)\right) F_{2,1}(g_1,g_2 s_2,g_2 s_2) = \left(\mu_2(s_2)\otimes\hat\mu_3(\hat s_3)\right) F\left(g_1^{-1} g_2 s_2\right),
\end{equation}
where $s_2,\hat s_3$ solve the equation $g_2 s_2 = g_3 \hat s_3$. The function $F$ is right-covariant with respect to the group $K$ and left-covariant with respect to $S_{d,p}$
\begin{equation}
    F(s g k) = F_{2,1}(e,s g k,s g k) = F_{2,1}(s^{-1},g k,g k) = \left(\mu_1(s)\otimes\mu_2(k^{-1})\otimes\hat\mu_3(k^{-1})\right) F(g)\ .
\end{equation}
Therefore, it can be regarded as a function on the double quotient
\begin{equation}\label{space-Y}
    Y = S_{d,p} \backslash G_{d,p}/ K = SO(p+1)\backslash SO(p+1,1) / SO(p)\ \times \ SO(q-1)\backslash SO(q) / SO(q-1)\ .
\end{equation}
Both direct factors were already analysed in previous subsections. The first one is two-dimensional and the second one one-dimensional. Cartan coordinates on the double coset are introduced by writing elements of $G_{d,p}$ as
\begin{equation}
    g = k_I e^{y K_1} e^{\lambda D} r^p\ r^{q-1}_l e^{\kappa M_{d-1,d}} r^{q-1}_r,
\end{equation}
with $r^p\in SO(p)$ and $r^{q-1}_{l,r}\in SO(q-1)$. The function $F$ is determined in terms of its restriction $\psi(\lambda,y,\kappa) = F(e^{y K_1} e^{\lambda D} e^{\kappa M_{d-1,d}})$ by
\begin{equation}
    F(g) = \left( \mu_1(k_I r_l^{q-1}) \otimes \mu_2 (r^p r^{q-1}_r)^{-1} \otimes \hat\mu_3 (r^p r^{q-1}_r)^{-1}\right) \psi(\lambda,y,\kappa)\ .
\end{equation}
The correlation function is related to $F$ by
\begin{equation}\label{G21-F}
    G_{2,1}(x_i) = \frac{1}{\Phi_1(x_1)\Phi_2(x_2)} F_{2,1}(g_d(x_1),g_d(x_2),g_p(\hat x_3)) = \frac{\mu_2(s_{23})\otimes\hat\mu_3(\hat s_{23})}{\Phi_1(x_1)\Phi_2(x_2)} F(g_d(x_1)^{-1}g_d(x_2) s_{23})\ .
\end{equation}
Here $s_{23}$ and $\hat s_{23}$ are given analogously to eq.\ \eqref{s-and-shat} . We can evaluate the argument of $F$ similarly as before using the Bruhat and Iwasawa decompositions. For the $SO(p+1)$-part
\begin{align*}
    & (g_d(x_1)^{-1}g_d(x_2) s_{23})_p = |x_{1\perp}|^{-D} e^{\hat x_{21}^a P_a} |x_{2\perp}|^D k_I\left(\frac{\hat x_{23}}{|x_{2\perp}|}\right)\\
    & = |x_{1\perp}|^{-D} e^{\hat x_{21}^a P_a} |x_{2\perp}|^D e^{-\frac{\hat x_{23}^a}{|x_{2\perp}|}P_a} \left(\frac{\hat x_{23}^2 + x_{2\perp}^2}{x_{2\perp}^2}\right)^D e^{\frac{\hat x_{23}^a}{|x_{2\perp}|}K_a} = e^{\frac{\hat x_{31}^a}{|x_{1\perp}|}P_a} \left(\frac{\hat x_{23}^2 + x_{1\perp}^2}{|x_{1\perp}| |x_{2\perp}|}\right)^D e^{\frac{\hat x_{23}^a}{|x_{2\perp}|}K_a}\\
    & = k_I\left(\frac{\hat x_{13}}{|x_{1\perp}|}\right) \left(\frac{\hat x_{13}^2 + x_{1\perp}^2}{x_{1\perp}^2}\right)^D e^{\frac{|x_{1\perp}|\hat x_{31}^a}{\hat x_{13}^2 + x_{1\perp}^2} K_a}\left(\frac{\hat x_{23}^2 + x_{2\perp}^2}{|x_{1\perp}| |x_{2\perp}|}\right)^D e^{\frac{\hat x_{23}^a}{|x_{1\perp}|}K_a}\\
    & = k_I\left(\frac{\hat x_{13}}{|x_{1\perp}|}\right) e^{\frac{1}{|x_{1\perp}|}\left(\hat x_{31} + \frac{\hat x_{13}^2 + x_{1\perp}^2}{\hat x_{23}^2 + x_{2\perp}^2}\hat x_{23}\right)\cdot K}\left(\frac{|x_{1\perp}|(\hat x_{23}^2 + x_{2\perp}^2)}{|x_{2\perp}|(\hat x_{13}^2 + x_{1\perp}^2)}\right)^D\ .
\end{align*}
We read off the coordinate $y$
\begin{align*}
    y & = \frac{1}{|x_{1\perp}|(\hat x_{23}^2 + x_{2\perp}^2)}\sqrt{\hat x_{12}^2 (\hat x_{13}^2 + x_{1\perp}^2)(\hat x_{23}^2+  x_{2\perp}^2) + (\hat x_{13}^2 x_{2\perp}^2 - \hat x_{23}^2 x_{1\perp}^2)(\hat x_{23}^2 + x_{2\perp}^2 - \hat x_{13}^2 - x_{1\perp}^2)}\\
    & = \frac{1}{|x_{1\perp}|(\hat x_{23}^2 + x_{2\perp}^2)}\sqrt{(\hat x_{12}^2 + x_{1\perp}^2 + x_{2\perp}^2) (\hat x_{13}^2 + x_{1\perp}^2)(\hat x_{23}^2+  x_{2\perp}^2) - x_{2\perp}^2 (\hat x_{13}^2 + x_{1\perp}^2)^2 - x_{1\perp}^2 (\hat x_{23}^2 + x_{2\perp}^2)^2 }\ .
\end{align*}
Therefore, the Cartan coordinates $(y,\lambda)$ are given by
\begin{equation}
    y = \sqrt{u_{12}^\bullet u_{3,12} - u_{3,12}^2 - 1}, \quad e^{\lambda} = u_{3,12}^{-1},
\end{equation}
where $u_{12}^\bullet$ and $u_{3,12}$ are cross ratios of \cite{Lauria:2017wav}
\begin{equation}\label{cross-ratios-Lauria-2}
    u_{12}^{\bullet} = \frac{\hat x_{12}^2 + x_{1\perp}^2 + x_{2\perp}^2}{|x_{1\perp}||x_{2\perp}|},\quad u_{3,12} = \frac{\hat x_{13}^2 + x_{1\perp}^2}{\hat x_{23}^2+ x_{2\perp}^2}\frac{|x_{2\perp}|}{|x_{1\perp}|}\ .
\end{equation}
The coordinate $\kappa$ was determined in eq.\ \eqref{kappa-coordinate} . These results give the relation between the correlator $G_{2,1}(x_i)$ and $\psi$ in the case the fields are scalars (it is possible to consider other representations as well, but we shall not do so)
\begin{equation}\label{bbd-correlator}
    G_{2,1}(x_i) = \frac{1}{|x_{1\perp}|^{\Delta_1} |x_{2\perp}|^{\Delta_2}} \left(\frac{\hat x_{23}^2 + x_{2\perp}^2}{|x_{2\perp}|}\right)^{-\Delta_{\hat 3}} \psi(\lambda,y,\kappa)\ .
\end{equation}
Eigenfunctions of the Laplacian correspond to eigenfunctions of the quadratic Casimir at the first insertion point $x_1$. Clearly, it is possible to repeat the whole argument with points $x_1$ and $x_2$ interchanged. This would lead to another representation of the correlator
\begin{equation}\label{bbd-correlator-tilde}
    G_{2,1}(x_i) = \frac{1}{|x_{1\perp}|^{\Delta_1} |x_{2\perp}|^{\Delta_2}} \left(\frac{\hat x_{13}^2 + x_{1\perp}^2}{|x_{1\perp}|}\right)^{-\Delta_{\hat 3}} \tilde\psi(\tilde\lambda,\tilde y,\kappa),
\end{equation}
with $\tilde y,\tilde\lambda$ obtained from $y,\lambda$ by swapping indices 1 and 2. One can parametrise the coset space $Y$ by $(y,\tilde y)$ and it is not hard to establish that $\tilde y = e^\lambda y\equiv x$. The Laplacian \eqref{Casimir-y-lambda}  in these coordinates reads
\begin{equation}
    C_2^{(2)} = (1+y^2)\partial_y^2 + \frac{x^2}{y^2}\partial_{x}^2 +\frac{2x}{y} \partial_y\partial_{x} + \left((p+1)y+\frac{p-1}{y}\right)\partial_y +\frac{(p-1)x}{y^2}\partial_{x}\ .
\end{equation}
If we performed and construction with $x_1$ and $x_2$ exchanged, the Laplacian would be given by the operator $C_2^{(1)}$ that is obtained from $C^{(2)}_2$ by exchanging $x$ and $y$. We have to remember that the prefactors multiplying $\psi$ and $\tilde\psi$ are different. Conformal blocks are therefore simultaneous eigenfunctions of $C_2^{(2)}$ and $(x/y)^{\Delta_{\hat3}}C_2^{(1)}(x/y)^{-\Delta_{\hat3}}$. These two operators are easily seen to commute. We will consider a more symmetric pair of differential operators
\begin{align}\label{eigenvalue-eqns}
    L_1 = \frac14\left(\frac{x}{y}\right)^{\frac{\Delta_{\hat3}}{2}}C_2^{(1)}\left(\frac{x}{y}\right)^{-\frac{\Delta_{\hat3}}{2}}, \quad  L_2 = \frac14\left(\frac{x}{y}\right)^{-\frac{\Delta_{\hat3}}{2}}C_2^{(2)}\left(\frac{x}{y}\right)^{\frac{\Delta_{\hat3}}{2}},
\end{align}
and solve an equivalent eigenvalue problem
\begin{equation}
    L_1 f(x,y) = \frac14\hat\Delta (\hat\Delta - p) f(x,y), \quad L_2 f(x,y) = \frac14\hat\Delta' (\hat\Delta' - p) f(x,y)\ .
\end{equation}
To proceed, let us introduce variables $v_1 = -x^{-2}$, $v_2 = -y^{-2}$. Then the two operators can be written as
\begin{align*}
    L_1 = v_1 D_{v_1 v_2}(0,\frac{2-p}{2},\frac{\Delta_{\hat3}-p+2}{2}) + \frac{\frac{\Delta_{\hat3}}{2}\left(\frac{\Delta_{\hat3}}{2}-p\right)}{4}, \ L_2 =  v_2 D_{v_2 v_1}(0,\frac{2-p}{2},\frac{\Delta_{\hat3}-p+2}{2}) + \frac{\frac{\Delta_{\hat3}}{2}\left(\frac{\Delta_{\hat3}}{2}-p\right)}{4},
\end{align*}
where $D_{xy}(a,b,c)$ is defined as
\begin{equation}
    D_{xy}(a,b,c) = x(1-x)\partial_x^2 - 2 x y \partial_x \partial_y - y^2 \partial_y^2 + (c-(a+b+1)x)\partial_x - (a+b+1)y\partial_y - ab\ .
\end{equation}
The significance of this operator is that it appears in connection with Appell's hypergeometric function $F_4$. Namely, this function carries four labels $(a,b,c_1,c_2)$ and satisfies the system of differential equations
\begin{equation}
    D_{xy}(a,b,c_1) F_4(x,y) = D_{yx}(a,b,c_2) F_4(x,y) = 0\ .
\end{equation}
Our equations are not quite in the form of the Appell's system, but they become so once we introduce $f(v_1,v_2) = v_1^{\frac{\hat\Delta}{2}-\frac{\Delta_{\hat3}}{4}} v_2^{\frac{\hat\Delta'}{2}-\frac{\Delta_{\hat3}}{4}} F(v_1,v_2)$. Then, using formulas from the appendix C, the eigenvalue equations \eqref{eigenvalue-eqns}  can be written in terms of $F$ as
\begin{align}
    & v_1 D_{v_1 v_2} \left(\frac{\hat\Delta+\hat\Delta'-\Delta_{\hat3}}{2},\frac{\hat\Delta+\hat\Delta'-\Delta_{\hat3}+2-p}{2},\hat\Delta-\frac{p}{2}+1\right) F = 0,\\
    & v_2 D_{v_2 v_1} \left(\frac{\hat\Delta+\hat\Delta'-\Delta_{\hat3}}{2},\frac{\hat\Delta+\hat\Delta'-\Delta_{\hat3}+2-p}{2},\hat\Delta'-\frac{p}{2}+1\right) F = 0\ .
\end{align}
Therefore, the Appell function \cite{Erdelyi}
\begin{equation}\label{Appell-solution}
    F(v_1,v_2) = F_4\left(\frac{\hat\Delta+\hat\Delta'-\Delta_{\hat3}}{2},\frac{\hat\Delta+\hat\Delta'-\Delta_{\hat3}+2-p}{2},\hat\Delta-\frac{p}{2}+1,\hat\Delta'-\frac{p}{2}+1;v_1,v_2\right),
\end{equation}
solves the eigenvalue problem. There are three more independent solutions, all expressible in terms of Appell functions, but the one we have written has the correct boundary behaviour and we will see that it reproduces the result of \cite{Lauria:2020emq} in a special limit. Before doing that, let us give the final formula for Laplacian eigenfunctions that correspond to conformal blocks. They are labelled by three quantum numbers $(\hat\Delta,\hat\Delta',s)$ and read\footnote{The relation between $\Psi$ and $\psi$ is $\Psi = (v_1/v_2)^{\frac{\Delta_{\hat 3}}{4}}\psi$. We use $\Psi$ in the final formula as it gives the most symmetric form of blocks.}
\begin{equation}\label{bdd-blocks}
    \Psi_{\hat\Delta,\hat\Delta',s}(v_1,v_2,\kappa) = v_1^{\frac{\hat\Delta}{2}-\frac{\Delta_{\hat3}}{4}} v_2^{\frac{\hat\Delta'}{2}-\frac{\Delta_{\hat3}}{4}} F(v_1,v_2) \ C^{(q-2)/2}_s (\cos\kappa)\ .
\end{equation}
The authors of \cite{Lauria:2020emq} consider the three-point function of two bulk and one defect field in the limit $\hat x_3\xrightarrow{}\infty$ and in the special configuration $x_{1\perp} = x_{2\perp}$. In such a configuration, $v_1=v_2$ and there are two independent cross ratios, $\varphi = \kappa$ and $\hat\chi=-v_1^{-1}$. The correlator in \cite{Lauria:2020emq} reads
\begin{equation}
    \hat x_3^{2\Delta_{\hat 3}}\langle \mathcal{O}_1(x_1) \mathcal{O}_2(x_2)\mathcal{\hat O}_3(\hat x_3) \rangle \sim \frac{e^{-i s_1\varphi_1}}{|x_{1\perp}|^{\Delta_1 + \Delta_2 - \Delta_{\hat3}}} \sum\mathcal{F}^{\mathcal{\hat O}_3}_{\mathcal{\hat O}\mathcal{\hat O}'}(\hat\chi)F(\varphi)\ .
\end{equation}
The conformal blocks factorise in $\hat\chi$ and $\varphi$ in the usual way and the transverse parts agree with ours by the same calculation as in the previous sections. Let us focus therefore on {\it longitudinal} parts, which are given in \cite{Lauria:2020emq} by
\begin{align}
    & \mathcal{F}^{\mathcal{\hat O}_3}_{\mathcal{\hat O}\mathcal{\hat O}'}(\hat\chi) = \hat\chi^{-\frac12(\hat\Delta+\hat\Delta'-\Delta_{\hat3})}\ _4F_3\Big(\frac{\hat\Delta+\hat\Delta'-p+1}{2},\frac{\hat\Delta+\hat\Delta'-p+2}{2},\\
    &\frac{\hat\Delta+\hat\Delta'-\Delta_{\hat3}}{2},\frac{\hat\Delta+\hat\Delta'-\Delta_{\hat3}-p+2}{2};\hat\Delta-\frac{p}{2}+1,\hat\Delta'-\frac{p}{2}+1,\hat\Delta+\hat\Delta'-p+1;-\frac{4}{\hat\chi}\Big)\ .
\end{align}
We can rewrite this using an identity due to Burchnall as
\begin{equation*}
     \mathcal{F}^{\mathcal{\hat O}_3}_{\mathcal{\hat O}\mathcal{\hat O}'}(\hat\chi) = \hat\chi^{-\frac12(\hat\Delta+\hat\Delta'-\Delta_{\hat3})}\ F_4\Big(\frac{\hat\Delta+\hat\Delta'-\Delta_{\hat3}}{2},\frac{\hat\Delta+\hat\Delta'-\Delta_{\hat3}-p+2}{2};\hat\Delta-\frac{p}{2}+1,\hat\Delta'-\frac{p}{2}+1;-\frac{1}{\hat\chi},-\frac{1}{\hat\chi}\Big)\ .
\end{equation*}
The prefactor in the correlation function that multiplies $\mathcal{F}^{\mathcal{\hat O}_3}_{\mathcal{\hat O}\mathcal{\hat O}'}(\hat\chi)$ is the same as the prefactor of $f$, so we need to show that $f(v_1,v_1)$ and $\mathcal{F}^{\mathcal{\hat O}_3}_{\mathcal{\hat O}\mathcal{\hat O}'}(\hat\chi)$ agree, up to a multiplicative constant. But one readily observes that
\begin{equation}
    \mathcal{F}^{\mathcal{\hat O}_3}_{\mathcal{\hat O}\mathcal{\hat O}'}(\hat\chi) = (-1)^{-\frac12(\hat\Delta+\hat\Delta'-\Delta_{\hat3})} f(v_1,v_1)\ .
\end{equation}
Therefore, the blocks from \cite{Lauria:2020emq} follow from those written in eq.\ \eqref{bbd-blocks}. The same is true for blocks of bulk-bulk-defect three point functions that were found in \cite{Behan:2020nsf}. This is shown the appendix D.

\section{Discussion and Outlook}

In this work we have extended the group theoretic approach to conformal blocks to the defect channel of  
correlation functions with any number $m$ and $n$ of bulk and defect field insertions, respectively. In
the most basic formulation, we obtained a new representation of correlation functions in defect conformal
field theory through covariant functions on $n+m$ copies of the defect conformal group. This representation was
then reduced to one that involves only $\textit{max}(m,n)-1$ copies of $G_{d,p}$ by forming
pairs of bulk and defect field insertions as  well as implementing conformal invariance. The details
for this representation were worked out for the case when the bulk fields are scalars. We then applied our lift to the construction
of correlation functions with a small number of field insertions. Once the setup allows for non-trivial
cross ratios, i.e. when there is more than one bulk- and one defect-field insertion, correlation functions
can be expanded into blocks. We characterised these as eigenfunctions of conformal Laplace-Beltrami
operators and constructed them explicitly for three different cases. The blocks for $m=2$ bulk fields and
$n=1$ defect fields, which we expressed in terms of Appell's hypergeometric function $F_4$, were not
known before, except along a lower dimensional submanifold in the space of cross ratios.
\smallskip

There is a number of natural extensions of the present work that should be pursued in the future. To begin
with, let us mention correlation functions of two bulk and two defect fields. Using the constructions we
explained above these correlators can still be lifted to functions on a single copy of the defect
conformal group. It would certainly be interesting to work out the Casimir equations and to construct
the associated conformal blocks. The setup admits five cross ratios, at least for sufficiently generic
dimensions $d,p$, and additional quantum numbers that label the weight and spin of an intermediate
defect field. One nice aspect of these correlation functions is that they give rise to non-trivial
(defect channel) crossing symmetry equations. The two sides of these equations are related by a
simple exchange of the two bulk fields. It would be very interesting to study these
constraints with either numerical or analytical bootstrap techniques.  

Going to even higher number of points, the number or cross ratios keeps growing. Even though such
correlators ultimately can no longer be lifted to a single copy of $G_{d,p}$, much of the group theoretic
technology remains at our disposal. An interesting direction would be to investigate what kind of
special functions appear as one increases the number of bulk and defect field insertions, and how
these depend on the dimensions $d$ and $p$. For line defects, correlators with multiple insertions
of defect fields (in addition to a few bulk fields) should be accessible  with more or less
know techniques, see \cite{Rosenhaus:2018zqn} for a related study of multi-point blocks in
1-dimensional bulk theories. More generally, one may expect all such multi-point correlators to
be wave functions of some integrable systems, in a similar way as multi-point correlators
in ordinary conformal field theory were recently identified as wave functions of Gaudin integrable
models \cite{Buric:2020dyz}, extending related observations in \cite{Isachenkov:2016gim,
Isachenkov:2017qgn,Isachenkov:2018pef}. The relation of defect channel blocks with integrable
quantum systems remains an interesting area to develop.
  
A more specific question in this context arises from the observation that Appell's functions,
though of type $F_2$, have appeared in the literature on conformal blocks, as so called
{\it multichannel blocks} for null heptagon Wilson loops, \cite{Sever:2011pc}. The framework
there is not entirely dissimilar from ours, as multichannel blocks depend non-trivially on
two variables and diagonalise two commuting quadratic Casimirs. On the other hand, ordinary
four-point blocks satisfy a different kind of system - they diagonalise quadratic and fourth
order Casimirs constructed from the same set of generators. It would be interesting to explore
if there is a deeper connection between calculations of multichannel blocks and the ones
considered here, perhaps in the framework of Gaudin models, \cite{Buric:2020dyz}. For example,
the system responsible for the integrability of five-point conformal blocks consists of five
operators, two of second order and three of fourth order. In a certain lightcone-like limit,
the system can be reduced to a three-variable problem with two quadratic and one quartic
operator. The two quadratic operators bear close resemblance to the ones in
\cite{Sever:2011pc}.
\smallskip

In this work we have restricted ourselves to the discussion of scalar fields, most notably when we constructed
the geometric lift of bulk fields to the defect conformal group in section 2.2. Following the general strategy
we outlined above it is essentially clear how to remove this restriction. After lifting spinning correlators to
the defect conformal group one could then construct and solve the associated matrix valued Casimir equations,
see \cite{Schomerus:2016epl,Schomerus:2017eny} for the corresponding analysis in bulk conformal field theory. For
defect channel blocks such as the ones we discussed here, it should be possible without much difficulty to
adapt the weight-shifting technology of \cite{Karateev:2017jgd} and thereby to solve the spinning Casimir
equations by acting on scalar blocks with appropriate weight shifting operators.

The study of defects becomes particularly rich in the context of superconformal theories. These can be
studied with several non-perturbative techniques including the conformal bootstrap, see e.g. \cite{Liendo:2016ymz,
Bianchi:2018zpb,DiPietro:2019hqe,Gimenez-Grau:2019hez,Bianchi:2019sxz,Gimenez-Grau:2020jvf}. Assuming the successful extension of our  
work to spinning defect correlators, see previous paragraph, it seems likely that the approach to superconformal
blocks we developed in \cite{Buric:2019rms,Buric:2020buk,Buric:2020qzp} for bulk four-point functions can be extended to theories with defects. We
plan to address this construction in the future.
\bigskip

{\bf Acknowledgements:} We wish to thank Aleix Gimenez-Grau, Edoardo Lauria, Pedro Liendo, Lorenzo Di Pietro, Evgeny Sobko and
Philine van Vliet for comments and fruitful discussion and acknowledge the support by the Deutsche Forschungsgemeinschaft (DFG,
German Research Foundation) under Germany's Excellence Strategy { EXC 2121 ,,Quantum Universe" {
390833306.}}

\appendix

\section{Conformal Group and Its Vector Representation}

Here we spell out our conventions for the conformal group $SO(d+1,1)$ and its Lie algebra and state some identities valid in the vector representation. The non-vanishing Lie brackets in $\mathfrak{so}(d+1,1)$ read
\begin{align}
    & [M_{\mu\nu},P_\rho] = \delta_{\nu\rho} P_\mu - \delta_{\mu\rho} P_\nu,\quad [M_{\mu\nu},K_\rho] = \delta_{\nu\rho} K_\mu - \delta_{\mu\rho} K_\nu,\\
    & [M_{\mu\nu},M_{\rho\sigma}] = \delta_{\nu\rho} M_{\mu\sigma} - \delta_{\mu\rho} M_{\nu\sigma} + \delta_{\nu\sigma} M_{\rho\mu} - \delta_{\mu\sigma} M_{\rho\nu},\\
    & [D,P_\mu] = P_\mu,\quad [D,K_\mu]=-K_\mu,\quad [K_\mu,P_\nu] = 2(M_{\mu\nu} - \delta_{\mu\nu}D)\ .
\end{align}
In the Lorentz-like notation, we write the generators as $\{L_{\alpha\beta}\}$, $\alpha,\beta = 0,1,...,d$. These obey the relations
\begin{equation}
    [L_{\alpha\beta},L_{\gamma\delta}] = \eta_{\beta\gamma} L_{\alpha\delta} - \eta_{\alpha\gamma} L_{\beta\delta} + \eta_{\beta\delta} L_{\gamma\alpha} - \eta_{\alpha\delta} L_{\gamma\beta},
\end{equation}
where $\eta$ is the mostly-positive Minkowski metric. The relation between conformal and Lorentz generators reads
\begin{equation}
    L_{01} = D, \quad L_{0\mu} = \frac12(P_\mu + K_\mu), \quad L_{1\mu} = \frac12(P_\mu - K_\mu), \quad L_{\mu\nu} = M_{\mu\nu}\ .
\end{equation}
The quadratic Casimir and its value in the representation $(\Delta,l)$ of $SO(d+1,1)$ are given by
\begin{equation}
    C_2 = -\frac12 L^{\alpha\beta} L_{\alpha\beta} = D^2 +\frac12 \{P_\mu,K^\mu\} -\frac12 M^{\mu\nu} M_{\mu\nu}, \quad C_2(\Delta,l) = \Delta(\Delta-d) + l(l+d-2)\ .
\end{equation}
In the $d+2$-dimensional vector representation, the Lorentz generators are
\begin{equation}
    L_{\alpha\beta} = \eta_{\alpha\gamma} E_{\gamma\beta} - \eta_{\beta\gamma} E_{\gamma\alpha},
\end{equation}
where $ (E_{\alpha\beta})_{ij} = \delta_{\alpha i} \delta_{\beta j}$. Thus in particular
\begin{equation}
    P_\mu = E_{1\mu} - E_{\mu1} - E_{0\mu} - E_{\mu0}, \quad K_\mu = - E_{1\mu} + E_{\mu1} - E_{0\mu} - E_{\mu0}\ .
\end{equation}
We will write matrices in the vector representation in block form. For example
\begin{equation}
    x^\mu P_\mu = \begin{pmatrix}
    0 & 0 & -x^T\\
    0 & 0 & x^T\\
    -x & -x & 0
    \end{pmatrix}, \quad x^\mu K_\mu = \begin{pmatrix}
    0 & 0 & -x^T\\
    0 & 0 & -x^T\\
    -x & x & 0
    \end{pmatrix}\ .
\end{equation}
The matrices representing translations and special conformal transformation are easily found using nilpotency of $P_\mu$ and $K_\mu$. Namely $(x^\mu P_\mu)^3 = (x^\mu K_\mu)^3 = 0$ and
\begin{align}
    & x^\mu x^\nu P_\mu P_\nu = x^\mu x_\mu(E_{00} - E_{11} + E_{01} - E_{10}), \quad x^\mu x^\nu K_\mu K_\nu = x^\mu x_\mu(E_{00} - E_{11} - E_{01} + E_{10})\ .
\end{align}
Therefore
\begin{equation}
    e^{x^\mu P_\mu} = \begin{pmatrix}
    1+\frac12 x^2 & \frac12 x^2 & -x^T\\
    -\frac12 x^2 & 1 - \frac12 x^2 & x^T\\
    -x & -x & 1
    \end{pmatrix}, \quad e^{x^\mu K_\mu} = \begin{pmatrix}
    1+\frac12 x^2 & -\frac12 x^2 & -x^T\\
    \frac12 x^2 & 1 - \frac12 x^2 & -x^T\\
    -x & x & 1
    \end{pmatrix}\ .
\end{equation}
The dilations are represented as
\begin{equation}
     D = - E_{01} - E_{10},\quad e^{\lambda D} = \begin{pmatrix}
     \cosh\lambda & -\sinh\lambda & 0\\
     -\sinh\lambda & \cosh\lambda & 0\\
     0 & 0 & 1
     \end{pmatrix}.
\end{equation}
The Iwasawa decomposition used in the main text follows from the matrix identity
\begin{equation}
    \begin{pmatrix}
    1+\frac12 x^2 & -\frac12 x^2 & -x^T\\
    \frac12 x^2 & 1 - \frac12 x^2 & -x^T\\
    -x & x & 1
    \end{pmatrix} = \begin{pmatrix}
    1+\frac12 y^2 & \frac12 y^2 & -y^T\\
    -\frac12 y^2 & 1 - \frac12 y^2 & y^T\\
    -y & -y & 1
    \end{pmatrix}\begin{pmatrix}
     \cosh\lambda & -\sinh\lambda & 0\\
     -\sinh\lambda & \cosh\lambda & 0\\
     0 & 0 & 1
     \end{pmatrix} \begin{pmatrix}
    1 & 0 & 0\\
    0 & \frac{1-x^2}{1+x^2} & \frac{-2x^\mu}{1+x^2}\\
    0 & \frac{2 x^\mu}{1+x^2} & \delta_{\mu\nu} - \frac{2x^\mu x^\nu}{1+x^2}
    \end{pmatrix},\nonumber
\end{equation}
where
\begin{equation}
    e^\lambda = \frac{1}{1+x^2}, \quad y^\mu = \frac{x^\mu}{1+x^2}\ .
\end{equation}
Let us also spell out the Weyl inversion in the vector representation
\begin{equation}
    w = e^{\pi\frac{K_d - P_d}{2}} = \text{diag}(1,-1,1,1,...,1,-1)\ .
\end{equation}

\section{Cross Ratios}

In this appendix we prove relations \eqref{lambda-coordinate}  and \eqref{kappa-coordinate} . The formula \eqref{lambda-coordinate}  is obtained by taking the $(1,1)$ matrix element in the vector representation of both sides of
\begin{equation}
    |x_{1\perp}|^{-D}  e^{x_{21}^a P_a} |x_{2\perp}|^D = r_l^{p+1}(x_i) e^{\lambda D} r_r^{p+1}(x_i)\ .
\end{equation}
For the coordinate defined in eq.\ \eqref{kappa-coordinate} consider the space $\mathbb{R}^q$ spanned by the
vectors $e_{p+1},...,e_{d}$. A direct calculation shows that
\begin{equation}
    e^{\varphi^i M_{id}} e_d = \left(\frac{\varphi^{p+1}\sin|\varphi|}{|\varphi|},...,\frac{\varphi^{d-1}\sin|\varphi|}{|\varphi|},\cos|\varphi|\right)^t, \quad |\varphi| = \left(\sum_{p+1}^{d-1} \varphi^i \varphi^i\right)^{\frac12}\ .
\end{equation}
By definition, this is the vector $x_\perp/|x_\perp|$ if the element $e^{\varphi^i M_{id}}$ is associated to $x$. Furthermore, the bottom right matrix element of $e^{-\varphi_1^i M_{id}} e^{\varphi_2^j M_{jd}}$ is seen to be
\begin{equation}
    \cos|\varphi_1|\cos|\varphi_2| + \frac{\sin|\varphi_1|\sin|\varphi_2|}{|\varphi_1| |\varphi_2|}\sum_{j=p+1}^{d-1}\varphi_1^j\varphi_2^j = \frac{x_1^i x_2^i}{|x_{1\perp}| |x_{2\perp|}}\ .
\end{equation}
This is compared with the bottom right element of the matrix $r_l^{q-1} e^{\kappa M_{d-1,d}} r_r^{q-1}$, which is $\cos\kappa$. Thus, eq.\ \eqref{kappa-coordinate}  follows.

\section{Hypergeometric and Appell's Functions}

In this appendix we collect some properties of Gauss' and Appell's hypergeometric functions. These are used in section 4 to arrive at conformal blocks. The hypergeometric differential operator depends on three parameters $a,b,c$ and we can write it as
\begin{equation}
    H(a,b,c,x,\partial_x) = x(1-x)\partial_x^2 + (c - (a+b+1)x) \partial_x - ab = (x\partial_x + c)\partial_x - (x\partial_x + a)(x\partial_x + b)\ .
\end{equation}
The hypergeometric differential equation $H(a,b,c,x,\partial_x)f=0$ has two independent solutions near the origin of the complex $x$-plane
\begin{equation}
    f_1 =\ _2F_1(a,b,c,x), \quad f_2 =\ x^{1-c}\ _2F_1(1+a-c,1+b-c,2-c,x)\ .
\end{equation}
To solve the eigenvalue problem $H(a,b,c,x,\partial_x)f=\lambda f$, one observes that it takes again the form of a hypergeometric equation with parameters $a',b',c$ such that $a'b'=ab+\lambda$ and $a'+b' = a+b$.

In the analysis of conformal blocks for four-point functions, a significant role is played by the differential operator $D_x^{a,b,c} = x  H(a,b,c,x,\partial_x)$, that was extensively studied by Dolan and Osborn, \cite{Dolan:2003hv,Dolan:2011dv}. Now the eigenvalue problem $D_x^{a,b,c} f = \lambda (\lambda + c - 1) f$ has independent solutions
\begin{equation}
    f_1 = x^\lambda\ _2F_1(a+\lambda,b+\lambda,c+2\lambda,x), \quad f_2 = x^{1-c-\lambda}\ _2F_1(1+a-c-\lambda,1+b-c-\lambda,2-c-2\lambda,x)\ .
\end{equation}
This follows from the identity
\begin{equation}
    x^{-\lambda} H(a,b,c,x,\partial_x) x^\lambda = H(a+\lambda,b+\lambda,c+2\lambda,x,\partial_x) + \frac{\lambda(\lambda+c-1)}{x}\ .
\end{equation}
Thus indeed
\begin{align}
    & D_x^{a,b,c} x^\lambda\ _2F_1(a+\lambda,b+\lambda,c+2\lambda,x) = x^{\lambda+1}\left(H(a+\lambda,b+\lambda,c+2\lambda,x,\partial_x) + \frac{\lambda(\lambda+c-1)}{x}\right)\\
    & _2F_1(a+\lambda,b+\lambda,c+2\lambda,x) = \lambda(\lambda+c-1)x^\lambda\ _2F_1(a+\lambda,b+\lambda,c+2\lambda,x)\ .
\end{align}
Appell's, or more generally Horn's differential equations are obtained by promoting the parameters $a,b,c$ to commuting operators in a variable $y$, \cite{symbolic}. Concretely, for the Appell's function $F_4$ we set
\begin{align}
    H_1 = H(a + y \partial_y, b + y \partial_y, c_1, x, \partial_x), \quad H_2 = H(a + x \partial_x, b + x \partial_x, c_2, y, \partial_y),
\end{align}
The associated system of equations reads $H_1 f(x,y) = H_2 f(x,y) = 0$. There are four independent solutions around the origin. We write only one of them
\begin{equation*}
    F_4(a,b,c_1,c_2,x,y) =\ _2F_1 (a+y\partial_y,b+y\partial_y,c_1,x)\ _2F_1(a,b,c_2,y) =\ _2F_1 (a+x\partial_x,b+x\partial_x,c_2,y)\ _2F_1(a,b,c_1,x)\ .
\end{equation*}
It is clear that from the first representation of $F_4$ that it solves the equation $H_1 F_4 =0$ and similarly from the second that it solves $H_2 F_4 = 0$. What is non-trivial is that the two representations give the same function. Similarly as in one-variable case we have
\begin{align}
    & x^{-\lambda} H_1 x^\lambda = H(a+\lambda+y\partial_y,b+\lambda+y\partial_y,c_1 + 2\lambda,x) + \frac{\lambda(\lambda + c_1 - 1)}{x}\ .
\end{align}
Further, one can readily verify
\begin{align}
     y^{-\mu} H_1 y^\mu = H(a+\mu+y\partial_y,b+\mu+y\partial_y,c_1,x)\ .
\end{align}
When combined, these two relations lead to
\begin{align}
    & y^{-\mu}x^{-\lambda} H_1 x^\lambda y^\mu = H(a+\lambda+\mu+y\partial_y,b+\lambda+\mu+y\partial_y,c_1 + 2\lambda,x) + \frac{\lambda(\lambda + c_1 - 1)}{x} \ .
\end{align}
Analogous statements hold for $H_2$. These are used to in section 4 to derive the system of equations for $F$ in terms of the one for $f$.

\section{Comparison with conformal blocks from \cite{Behan:2020nsf}}

In this appendix, we show how our conformal blocks for the correlator $G_{2,1}(x_i)$ reproduce as a special case the blocks from \cite{Behan:2020nsf}. The partial waves from that paper appear in the analysis of a free scalar bulk field in the presence of a boundary. Coordinates $\xi$ and $\zeta$ used in \cite{Behan:2020nsf} are in our notation $\xi=u_{12}^\bullet-2$ and $\zeta = u_{3,12}^{-1}$. Therefore, the $w_\pm$ of \cite{Behan:2020nsf} read
\begin{equation}
    w_\pm = \left(\frac1x \pm \frac1y\right)^2\ .
\end{equation}
To restrict our analysis to the boundary setup, notice first that for $q=1$, the second factor in the cross ratio space $(\ref{space-Y})$ becomes trivial, so our blocks become
\begin{equation}\label{psi-ij}
    \Psi_{\hat\Delta,\hat\Delta'}(v_1,v_2) = v_1^{\frac{\hat\Delta}{2}-\frac{\Delta_{\hat3}}{4}} v_2^{\frac{\hat\Delta'}{2}-\frac{\Delta_{\hat3}}{4}} F(v_1,v_2)\ .
\end{equation}
Furthermore, the fact that the bulk field is free means that the fields that propagate in the bulk-defect OPE are highly restricted. Dimensions $\hat\Delta$, $\hat\Delta'$ of intermediate fields can only assume two values that we denote by
\begin{equation}
    \hat\Delta_1 = \frac{d}{2} - 1, \quad \hat\Delta_2 = \frac{d}{2}\ .
\end{equation}
Let us write $F^{ij}$ for Appell's functions $(\ref{Appell-solution})$ with $\hat\Delta=\hat\Delta_i$ and $\hat\Delta'=\hat\Delta_j$. The corresponding conformal blocks $(\ref{psi-ij})$ are denoted $\Psi_{ij}$. We have
\begin{align}
    & F^{11} = F_4\Big(\frac{d-2-\Delta_{\hat3}}{2},\frac{1-\Delta_{\hat3}}{2};\frac12,\frac12;v_i\Big)\nonumber\\
    & =\frac12 \left(\, _2F_1\left(\frac{1-\Delta_{\hat3}}{2},\frac{d-2-\Delta_{\hat3}}{2};\frac12;\left(\sqrt{v_1}-\sqrt{v_2}\right)^2\right)+\, _2F_1\left(\frac{1-\Delta_{\hat3}}{2},\frac{d-2-\Delta_{\hat3}}{2};\frac12;\left(\sqrt{v_1}+\sqrt{v_2}\right)^2\right)\right)\nonumber\\
    & = \frac12\left(\, _2F_1\left(\frac{1-\Delta_{\hat3}}{2},\frac{d-2-\Delta_{\hat3}}{2};\frac12;-w_-\right)+(w_-\leftrightarrow w_+)\right)
\end{align}
In the first step, we expanded $F_4$ as a double series using its definition and rearranged the terms. Similarly
\begin{align}
    F^{22} & =F_4\Big(\frac{d-\Delta_{\hat3}}{2},\frac{3-\Delta_{\hat3}}{2};\frac32,\frac32;v_i\Big) \nonumber \\
    & =\frac{(v_1 v_2)^{-\frac12}}{2(\Delta_{\hat3}-1)(d-2-\Delta_{\hat3})}\left(\, _2F_1\left(\frac{1-\Delta_{\hat3}}{2},\frac{d-2-\Delta_{\hat3}}{2};\frac12;-w_-\right)-(w_-\leftrightarrow w_+)\right),\\
    F^{12} &= F_4\Big(\frac{d-1-\Delta_{\hat3}}{2},\frac{2-\Delta_{\hat3}}{2};\frac12,\frac32;v_i\Big)\nonumber\\
    & =\frac12 v_2^{-1/2} \left(i\sqrt{w_+}\, _2F_1\left(\frac{2-\Delta_{\hat3}}{2},\frac{d-1-\Delta_{\hat3}}{2};\frac32;-w_+\right)-(w_+\leftrightarrow w_-)\right),\\
    F^{21} &= F_4\Big(\frac{d-1-\Delta_{\hat3}}{2},\frac{2-\Delta_{\hat3}}{2};\frac32,\frac12;v_i\Big)\nonumber\\
    & =\frac12 v_1^{-1/2} \left(i\sqrt{w_+}\, _2F_1\left(\frac{2-\Delta_{\hat3}}{2},\frac{d-1-\Delta_{\hat3}}{2};\frac32;-w_+\right)+(w_+-\leftrightarrow w_-)\right)\ .
\end{align}
We therefore have
\begin{equation}
    \Psi_{11} = (v_1 v_2)^{\frac{d-2-\Delta_{\hat3}}{4}} \mathcal{F}^{11}_{\Delta_{\hat3},0}, \quad  \Psi_{22} = (v_1 v_2)^{\frac{d-2-\Delta_{\hat3}}{4}} \mathcal{F}^{22}_{\Delta_{\hat3},0}, \quad \Psi_{12} + \Psi_{21} = i(v_1 v_2)^{\frac{d-2-\Delta_{\hat3}}{4}} \mathcal{F}^{12}_{\Delta_{\hat3},0}\ .
\end{equation}
Using that in the limit $\hat x_3\xrightarrow{}\infty$, we have $v_i = - x^2_{i^\vee\perp}$, the last equation together with e.q.\ \eqref{bbd-correlator} gives the formula (B.6) from \cite{Behan:2020nsf} with $l=0$.

\end{document}